\documentclass[%
 reprint,
 amsmath,amssymb,
 aps,
prb,
 longbibliography
]{revtex4-1}

\usepackage{graphicx}
\usepackage{dcolumn}
\usepackage{bm}
\usepackage{physics}
\usepackage{mleftright}
\usepackage{siunitx}
\usepackage{makecell}
\usepackage{hyperref}

\usepackage{ulem}
\usepackage[usenames]{xcolor}
\usepackage{epstopdf}

\definecolor{mygreen}{RGB}{0,204,102}

\DeclareMathAlphabet{\mathpzc}{OT1}{pzc}{m}{it}

\begin{document}

\preprint{APS/123-QED}

\title{Computational design of $f$-electron Kitaev magnets: honeycomb and hyperhoneycomb compounds $A_2$PrO$_3$ ($A=$ alkali metals)}

\author{Seong-Hoon Jang}
\author{Ryoya Sano}
\author{Yasuyuki Kato}
\author{Yukitoshi Motome}
\affiliation{
 Department of Applied Physics, The University of Tokyo, Tokyo 113-8656, Japan
}

\date{\today}

\begin{abstract}
The Kitaev spin model offers an exact quantum spin liquid in the ground state, which has stimulated exploration of its material realization over the last decade. Thus far, most of the candidates are found in $4d$- and $5d$-electron compounds, in which the low-spin $d^5$ electron configuration subject to strong spin-orbit coupling comprises a Kramers doublet with the effective angular momentum $j_{\rm eff}=1/2$ and gives rise to the bond-dependent anisotropic interactions in the Kitaev model. Here we theoretically investigate another candidates in $4f$-electron compounds with the $f^1$ electron configuration on both quasi-two-dimensional honeycomb and three-dimensional hyperhoneycomb structures, $A_2$PrO$_3$ with $A$=Li, Na, K, Rb, and Cs. Based on {\it ab initio} calculations, we show that the electronic structures of these compounds host a spin-orbital entangled Kramers doublet with $j_{\rm eff}=1/2$ in the $\Gamma_7$ state. By constructing the tight-binding Hamiltonian from maximally-localized Wannier functions and performing a perturbation expansion in the strong coupling limit, we find that the low-energy magnetic properties of $A_2$PrO$_3$ are well described by an effective spin model with the isotropic Heisenberg, anisotropic Kitaev, and symmetric off-diagonal interactions, dubbed the $J$-$K$-$\Gamma'$ model. The most remarkable feature is that the Kitaev interaction $K$ can be antiferromagnetic, in contrast to the ferromagnetic one in the $d^5$ candidates at hand. We show that the exchange interactions are systematically modulated by changing the $A$-site cations; while increasing the $A$-site ionic radii, $J$ is not largely modulated but $K$ is reduced and $\Gamma'$ is slightly increased. As a consequence, the compounds with $A$=Li and Na may have a dominant antiferromagnetic $K$, but $J$ dominates $K$ and $\Gamma'$ in the cases with $A$=Rb and Cs. We analyze the systematic changes by decomposing each interaction into the contributions from different perturbation processes. Also, by computing the ground states of the $J$-$K$-$\Gamma'$ model by using the exact diagonalization, we map out the systematic evolution of the model parameters in the phase diagram. Our results will stimulate material exploration of the antiferromagnetic Kitaev interaction in $f$-electron compounds, including the previously-synthesized honeycomb and hyperhoneycomb compounds, Na$_2$PrO$_3$.
\end{abstract}

\pacs{Valid PACS appear here}
\maketitle

\section{Introduction} 
\label{sec:introduction}

Electron correlation and spin-orbit coupling (SOC) are two crucial factors in the design of quantum materials. Beyond the conventional band theory for metals and insulators, the strong electron correlation may yield Mott insulators and anomalous metallic states~\cite{MO1968,IM1998}, which may endow high-temperature superconductivity. Meanwhile, the SOC entangles the orbital motion of electrons with the spin degree of freedom, leading to topological insulators~\cite{HA2010, QI2011} and topological semimetals~\cite{WE2016, BA2016A, CH2016A}. In recent years, it has been recognized that the synergy of the strong electron correlation and the SOC provides a fertile ground for yet another quantum states of matter, such as topological Mott insulators, Weyl semimetals, and axion insulators~\cite{WI2014}.

The quantum spin liquid (QSL) is one of such intriguing phases potentially induced by the electron correlation and SOC. It is a massively entangled quantum phase in which interacting localized magnetic moments are prevented from forming a magnetic long-range order by strong quantum fluctuations~\cite{AN1973, BA2010, ZH2017A, SA2017}. The fluctuating moments under the quantum entanglement can show a topological order~\cite{WE1991A, LE2006} and quantum number fractionalization into nonlocal quasiparticle excitations~\cite{SA1992, NA2008}. In particular, nonabelian quasiparticles, which obey neither Bose-Einstein nor Fermi-Dirac statistics, have attracted great interest from application to decoherence-free topological quantum computing~\cite{KI2003}. While the prototypical candidates for the QSLs have been explored in geometrically frustrated antiferromagnets lying on triangular-based lattice structures~\cite{LA2011, DI2013}, the spin-orbital entanglement by the SOC in the Mott insulators can offer another playground through the frustration between bond-dependent exchange interactions, dubbed compass-type interactions, even on unfrustrated lattice structures~\cite{NU2015}.

The Kitaev model is one of the pragmatic models with such exchange frustration. The model has bond-dependent Ising-type interactions on a honeycomb structure, whose Hamiltonian is given by~\cite{KI2006B} 
\begin{equation}
\mathpzc{H}= \sum_{\mu}\sum_{\langle i,i' \rangle_\mu}K^\mu S_i^\mu S_{i'}^\mu,
\label{eq:H_Kitaev}
\end{equation} 
where the summations are taken for the nearest-neighbor sites $i$ and $i'$ on the $\mu$ bonds ($\mu=x,y,z$ distinguishes the three types of bonds on the honeycomb structure); $K^\mu$ describes the coupling constant for the Ising-type interactions on the $\mu$ bonds, and $S_i^\mu$ represents the $\mu$ component of spin-1/2 operator at site $i$. As it is impossible to optimize the exchange energy on all the bonds simultaneously, the Kitaev model has severe frustration. Nevertheless, the ground state was exactly obtained as an exact QSL, whose excitations are described by fractional quasiparticles, itinerant Majorana fermions and localized $Z_2$ fluxes~\cite{KI2006B}.

While the original proposal by Kitaev was rather mathematical, Jackeli and Khaliullin pointed out the possible realization of the Kitaev model as a low-energy effective model for a certain series of oxides~\cite{JA2009}. In their theory, the effective spin-1/2 moments are given by the spin-orbital entangled Kramers doublet in the low-spin $d^5$ electron configuration under an octahedral crystal field (OCF) and strong SOC. These moments interact with each other via the Kitaev-type interactions predominantly when the conventional Heisenberg interactions are cancelled out by quantum interference between different perturbation processes via the ligands in edge-sharing $MX_6$ octahedra ($M$ and $X$ represent a transition metal cation and a ligand ion, respectively). Stimulated by this idea, material-oriented researches toward the Kitaev-type QSL have been done explosively over the last decade for the low-spin $4d^5$ and $5d^5$ electron compounds~\cite{TR2017, WI2017, HE2018, KN2019, TA2019, MO2019}, such as quasi-two-dimensional (2D) honeycomb magnets $A_2$IrO$_3$ ($A$=Li, Na)~\cite{SI2010, SI2012} and $\alpha$-RuCl$_3$~\cite{PL2014, KU2015}, three-dimensional (3D) hyperhoneycomb magnet $\beta$-Li$_2$IrO$_3$~\cite{TA2015}, and 3D stripy honeycomb magnet $\gamma$-Li$_2$IrO$_3$~\cite{MO2014B}. Among a lot of efforts to identify the nature of the Kitaev QSL in these candidates, a recent highlight is the observation of the half-quantized thermal Hall conductivity in $\alpha$-RuCl$_3$ as evidence of a gapped topological state of the Majorana fermions~\cite{KA2018B}. In addition, by extending the argument by Jackeli and Khaliullin, the high-spin $d^7$ electron systems have also been studied as another candidates with similar Kramers doublet~\cite{LI2018, SA2018, YA2019A, YA2019B, ZH2019}.

Recently, rare-earth materials, in which the strong SOC coexists with electron correlations, have attracted attention for materialization of the Kitaev-type interaction. For instance, Yb$^{3+}$-based compounds with $4f^{13}$ electron configurations were nominated~\cite{RA2018, LU2019}, and indeed, Kitaev QSL behavior was argued for YbCl$_3$, whose crystal structure is the same as $\alpha$-RuCl$_3$~\cite{XI2019}. Another promising candidate is found for the electron-hole counterpart, $4f^1$ electron configurations. In this category, Pr$^{4+}$-based materials are noteworthy, as several polymorphic structures of $A_2$PrO$_3$ ($A=$ alkali metals) hosting edge-sharing PrO$_6$ octahedra have been synthesized: for example, quasi-one-dimensional chain~\cite{WO1987, HI2006}, layered honeycomb~\cite{HI2006}, triangular~\cite{BR1977, PA1966}, and hyperhoneycomb structures~\cite{WO1988}. Theoretically, the authors proposed that, based on {\it ab initio} calculations, the magnetic properties of the quasi-2D honeycomb form of $A_2$PrO$_3$ with $A$=Li and Na are well described by the model with dominant antiferromagnetic (AFM) Kitaev interactions~\cite{JA2019}. This allows one to access unexplored parameter regions of the Kitaev QSLs, as the existing candidates with $4d$ and $5d$ electrons are believed to possess the ferromagnetic (FM) Kitaev interactions. The AFM Kitaev model has recently been captivated by its possibility of a field-induced exotic state that cannot be achieved for the FM Kitaev model~\cite{ZH2017B, GO2018, NA2018, RO2019, HI2019, PA2019}. Despite the intriguing possibility, the previous study in Ref.~\onlinecite{JA2019} was limited to the honeycomb materials with $A$=Li and Na. Given the various polymorhs, further studies are desired for the Pr-based materials. 

In this paper, we perform a systematic study of the electronic and magnetic properties of the Pr-based quasi-2D compounds $A_2$PrO$_3$ for the $A$-site substitution beyond the previous study~\cite{JA2019}. We also extend our analysis to the 3D hyperhoneycomb structure, which is realized in $\beta$-Na$_2$PrO$_3$~\cite{WO1988}. For the quasi-2D honeycomb cases, by {\it ab initio} calculations with structural optimization, we show that the $4f^1$ states under the strong SOC and the OCF are well approximated by the $\Gamma_7$ Kramers doublet with the effective angular momentum $j_{\rm eff}=1/2$ for all the $A$-site substitutions by alkali atoms (Li, Na, K, Rb, and Cs). We find that larger $A$-site ionic radii lead to not only a longer bond length between Pr cations but also larger trigonal distortions of PrO$_6$ octahedra, which bring about larger deviations from the ideal $\Gamma_7$ Kramers doublet. Based on the maximally-localized Wannier functions (MLWFs) obtained by the {\it ab initio} calculations, we construct multiorbital Hubbard models for these compounds. We study the low-energy magnetic properties of these models by deriving effective spin models in the strong coupling limit by a perturbation expansion in terms of direct $4f$-$4f$ and indirect $4f$-$2p$-$4f$ hoppings. We show that the effective spin models are described by three dominant exchange interactions: isotropic Heisenberg $J$, anisotropic Kitaev $K$, and symmetric off-diagonal $\Gamma'$. We find that the coupling constants change systematically by the $A$-site substitution; the increase in the $A$-site ionic radii suppresses the AFM $K$ and slightly increases $\Gamma'$, while it does not modulate the AFM $J$ substantially. As a consequence, the AFM $K$, which is dominant for $A$=Li and Na, becomes smaller than the AFM $J$ for $A$=K and Rb, and even changes the sign to be weakly FM for $A$=Cs. We also calculate the ground-state phase diagram for the $J$-$K$-$\Gamma'$ model by using the exact diagonalization of 24-site clusters, and discuss the systematic changes of the exchange coupling constants on the phase diagram. We find that the $A$=Li case is the most proximate to the AFM Kitaev QSL, and the increase of the $A$-site ionic radii shifts the system into the deep inside of the N\'eel ordered phase. We also perform similar analyses for the 3D hyperhoneycomb compound $\beta$-Na$_2$PrO$_3$. In this case, we bypass the structural optimization in the {\it ab initio} calculations by using the experimental lattice parameters. We show that the values of the exchange coupling constants for this compound are similar to those for the quasi-2D counterpart, suggesting that the compound provides a good platform for the 3D $J$-$K$-$\Gamma'$ model with the dominant AFM Kitaev coupling. 

The organization of the rest of this paper is as follows. In Sec.~\ref{sec:method}, we describe the details of our method: {\it ab initio} calculations of the electronic structures, construction of the multiorbital Hubbard model from the MLWF analysis, formation of the $j_{\rm eff}=1/2$ pseudospin in the $\Gamma_7$ doublet, and derivation of the effective pseudospin Hamiltonian by the perturbation expansion in the strong coupling limit. In Sec.~\ref{sec:honeycombmagnets}, we show the results for the quasi-2D honeycomb compounds $A_2$PrO$_3$. We discuss the systematic changes in the electronic band structure obtained by the {\it ab initio} calculations including optimized lattice structures, the tight-binding parameters obtained by the MLWF analysis, and the exchange coupling constants in the effective pseudospin Hamiltonian derived by the perturbation expansion for $A$-site substitution (Sec.~\ref{sec:structureoptimization}-\ref{sec:couplings}). In Sec.~\ref{sec:decomposition}, we analyze the results in detail by decomposing the contributions to each coupling constant into different perturbation processes. We also map out the systematic evolution on the magnetic ground-state phase diagram for the $J$-$K$-$\Gamma'$ model in Sec.~\ref{sec:magneticphase}. In Sec.~\ref{sec:hyperhoneycombmagnets}, we present similar analyses for the experimentally-synthesized 3D hyperhoneycomb compound $\beta$-Na$_2$PrO$_3$~\cite{WO1988}. Section~\ref{sec:summary} is devoted to the summary. In Appendices~\ref{app:multiplets} and \ref{app:multiplets2}, we show the details of the multiplets given for the $4f^1$ and the $4f^2$ electron configurations, respectively.

\section{Method} 
\label{sec:method}

In this section, we introduce the theoretical methods used in this paper. In Sec.~\ref{subsec:ab_initio}, we present the details of the {\it ab initio} calculations and the MLWF analysis. In Sec.~\ref{subsec:Multiorbital}, we introduce the Hamiltonian for the multiorbital Hubbard model, whose parameters for the electron hopping are obtained by the MLWF analysis. In Sec.~\ref{subsec:KramersDoublet}, we show that the atomic electronic state for the $4f^1$ electron configuration under the strong SOC and the OCF yields the $\Gamma_7$ Kramers doublet with the effective angular momentum $j_{\rm eff}=1/2$. In Sec.~\ref{subsec:perturbation}, we introduce the perturbation scheme in the strong coupling limit to derive the low-energy effective Hamiltonian for the $j_{\rm eff}=1/2$ pseudospins.

\subsection{{\it ab initio} calculation of electronic structures}
\label{subsec:ab_initio}

In the {\it ab initio} calculations, we study the electronic structures of the quasi-2D layered honeycomb compounds $A_2$PrO$_3$ with $A$=K, Rb, and Cs, and the 3D hyperhoneycomb compound $\beta$-Na$_2$PrO$_3$. For the former honeycomb cases, the results for $A$=Li and Na are available in Ref.~\onlinecite{JA2019}. For the latter hyperhoneycomb case, we focus on the Na case, as the structural data is available only for the Na compound and the structural optimization is computationally expensive for other $A$-site ions because of the large number of atoms in the unit cell. All the {\it ab initio} calculations are performed by using \texttt{Quantum ESPRESSO}~\cite{GI2017}.

In the calculations for the honeycomb compounds, we adopt the pseudopotentials of non-relativistic norm-conserving Hartwigesen-Goedecker-Hutter type~\cite{HA1998} for the $A$-site cations ($A$=K, Rb, and Cs) and the O ions, while the full-relativistic ultrasoft projector-augmented-wave-method Perdew-Zunger type~\cite{PE1981, BL1994} for the Pr cations~\cite{[Note that different types of the pseudopotentials were adopted in the previous study for $A$=Li and Na in Ref.~\onlinecite{JA2019} because of the technical reasons] DU0000}
.
We set the kinetic energy cutoff at 250~Ry. We perform the structural optimization starting from the structural parameters for Rb$_2$CeO$_3$ listed in \texttt{Materials Project}~\cite{JA2013B}. In the structural optimization, we set the criteria for the maximum crystal stress at 0.1~GPa. The remnant maximum atomic forces are less than 0.002~Ry/Bohr in the {\it ab}-plane and less than 0.0001~Ry/Bohr along the axis perpendicular to the plane. All the results for $A$=K, Rb, and Cs converge onto monoclinic structures with $C2/m$ symmetry as in the previous study for $A$=Li and Na~\cite{JA2019}. 

Meanwhile, in the calculations for the hyperhoneycomb compound $\beta$-Na$_2$PrO$_3$, we adopt the pseudopotentials of non-relativistic norm-conserving von Barth-Car type~\cite{BA1985} for Na, non-relativistic ultrasoft projector-augmented-wave-method Perdew-Zunger type~\cite{PE1981, BL1994} for Pr, and non-relativistic norm-conserving Hartwigesen-Goedecker-Hutter type~\cite{HA1998} for O. We use the experimental structure with $C2/c$ symmetry~\cite{WO1988} without further structural optimization. Using the electron hopping parameters from the MLWF analysis for the non-relativistic {\it ab initio} calculations, we construct the multiorbital Hubbard model by adding the SOC by hand; we take the SOC coefficient $\lambda$=120~meV, which was estimated for the quasi-2D honeycomb compound $\alpha$-Na$_2$PrO$_3$~\cite{JA2019} (see Sec.~\ref{sec:electronicstructure2}).

In both calculations, we compute the electronic band structures, the (projected) density of states, and the construction of MLWFs by using the Monkhorst-Pack grids~\cite{MO1976} of $4\times4\times4$ and $8\times8\times8$ ${\bf k}$-points determined from the primitive cells. We set the convergence threshold in the self-consistent field calculations at $1.0\times10^{-10}$~Ry. 
We construct the MLWFs by using \texttt{WANNIER90}~\cite{MO2014}. 

\subsection{Multiorbital Hamiltonian} 
\label{subsec:Multiorbital}

For both quasi-2D honeycomb and 3D hyperhoneycomb cases, we construct multiorbital Hubbard models for the $f$-orbital manifold on the basis of the {\it ab initio} results. The Hamiltonian is commonly composed of four terms as
\begin{equation}
\mathpzc{H}=\mathpzc{H}_{\textrm{SOC}}+\mathpzc{H}_{\textrm{CEF}}+\mathpzc{H}_{\textrm{int}}+\mathpzc{H}_{\textrm{hop}}.
\label{eq:H_multi}
\end{equation} 

The first term $\mathpzc{H}_{\textrm{SOC}}$ describes the effect of the SOC. 
The Hamiltonian is given by
\begin{equation}
\mathpzc{H}_{\textrm{SOC}} = \sum_i \mathpzc{H}_{\textrm{SOC},i},
\end{equation}
where
\begin{widetext}
\begin{equation}
\mathpzc{H}_{\textrm{SOC},i} = \frac{\lambda}{2}\sum_{m=-\ell}^{\ell}\sum_{\sigma}m\sigma \tilde{c}^{\dagger}_{im\sigma}\tilde{c}_{im\sigma} + \frac{\lambda}{2}\sum_{m=-\ell}^{\ell-1}\sqrt{\ell+m+1}\sqrt{\ell-m}(\tilde{c}^{\dagger}_{im+1-}\tilde{c}_{im+}+\tilde{c}^{\dagger}_{im+}\tilde{c}_{im+1-}),
\label{eq:H_SOC}
\end{equation} 
\end{widetext}
where $\lambda>0$ is the SOC coefficient, $\ell$ is the orbital quantum number taken as $\ell=3$ for the $f$-orbital manifold, and $m$ and $\sigma=\pm1$ denote the magnetic and spin quantum numbers, respectively; $\tilde{c}_{im\sigma}^\dagger$ and $\tilde{c}_{im\sigma}$ represent creation and annihilation operators of an electron with $m$ and $\sigma$ at site $i$ in the spherical harmonics basis, respectively. 

The second term in Eq.~(\ref{eq:H_multi}), $\mathpzc{H}_{\textrm{CEF}}$, describes the effect of the crystalline electric field. It is in general described by the rank-$r$ Stevens multipole operators $O_{rs}$ ($s=-r,-r+1,\cdots,r$) as
\begin{equation}
\mathpzc{H}_{\textrm{CEF}}=\sum_{r,s}B_{rs}O_{rs},
\label{eq:H_CEF}
\end{equation} 
where $B_{rs}$ denotes the coupling coefficient. In the present situation, the dominant contribution is the OCF from the oxygen ions octahedrally coordinated around the Pr$^{4+}$ cation, which we denote $\mathpzc{H}_{\textrm{OCF}}$. In the OCF, the only nonzero coefficients are $B_{44}=5B_{40}$ and $B_{64}=-21B_{60}$, and hence, $\mathpzc{H}_{\textrm{OCF}}$ is given by 
\begin{equation}
\mathpzc{H}_{\textrm{OCF}}=B_{40}(O_{40}+5O_{44})+B_{60}(O_{60}-21O_{64}).
\label{eq:H_CEFOh}
\end{equation} 
For Pr$^{4+}$-based materials with the OCF, $B_{40}$ and $B_{60}$ are positive and negative, respectively, and $B_{60}\simeq-0.004 B_{40}$~\cite{KE1985}. We take into account $\mathpzc{H}_{\textrm{OCF}}$ only in the following discussions in this section, while all other contributions are incorporated in Sec.~\ref{sec:result} by the MLWF analysis under realistic lattice structures.  

The third term in Eq.~(\ref{eq:H_multi}) describes the Coulomb interactions between $f$ electrons. The Hamiltonian is given by
\begin{widetext}
\begin{equation}
\mathpzc{H}_{\textrm{int}} = \sum_i \sum_{m_1,m_2,m_3,m_4} \sum_{\sigma_1,\sigma_2} \delta_{m_1+m_2,m_3+m_4} \sum_{k=0,2,4,6} F^kC^{(k)}(m_1,m_4)C^{(k)}(m_2,m_3) \tilde{c}^{\dagger}_{im_1\sigma_1}\tilde{c}^{\dagger}_{im_2\sigma_2}\tilde{c}_{im_3\sigma_2}\tilde{c}_{im_4\sigma_1},
\label{eq:H_int}
\end{equation}
\end{widetext}
where $F^k$ and $C^{(k)}$ denote the Slater-Condon parameters and the Guant coefficients, respectively ($k=0,2,4,6$); $\delta$ is the Kronecker delta. Here, the Slater-Condon parameters are related with the onsite Coulomb interaction $U$ and the Hund's-rule coupling $J_{\rm H}$ as~\cite{AN1993, AN1997} 
\begin{eqnarray}
U&=&F^0,
\label{eq:U}
\\
J_{\textrm H}&=&\frac{1}{6435}\left( 286F^2+195F^4+250F^6 \right).
\label{eq:JH}
\end{eqnarray}

The fourth term in Eq.~(\ref{eq:H_multi}) describes the kinetic energy as
\begin{equation}
\mathpzc{H}_{\textrm{hop}} = \sum_{\mu}\sum_{\langle i,i' \rangle_\mu}\mathpzc{H}^{ \left( \mu \right) }_{\textrm{hop},ii'} ,
\label{eq:Hhop} 
\end{equation}
where $\mathpzc{H}^{ \left( \mu \right) }_{\textrm{hop},ii'}$ denotes the electron hopping between nearest-neighbor sites $i$ and $i'$ on the $\mu$ bond (one of the three types of bonds on the tricoordinate structure, labeled as $\mu=x$, $y$, and $z$) as
\begin{equation}
\mathpzc{H}^{ \left( \mu \right) }_{\textrm{hop},ii'} = \sum_{u,v}\sum_{\sigma=\pm}(\tilde{t}_{iu,i'v,\sigma}c^{\dagger}_{iu\sigma}c_{i'v\sigma}+\textrm{h.c.}).
\label{eq:Hhopij}
\end{equation}
Here, $\tilde{t}_{iu,i'v,\sigma}$ denotes the effective transfer integral between $f$ orbital $u$ at site $i$ and $f$ orbital $v$ at site $i'$ for spin $\sigma$ ($u$ and $v$ represent the seven types of $4f$ orbitals in the cubic harmonic basis, $\xi$, $\eta$, $\zeta$, $A$, $\alpha$, $\beta$, and $\gamma$~\cite{[As regards the linear transformation between spherical and cubic harmonic bases{\rm ,} refer to  ]TA1980}), which includes contributions from both direct $4f$-$4f$ and indirect $4f$-$2p$-$4f$ hopping processes; $c^{\dagger}_{iu\sigma}$ and $c_{iu\sigma}$ represent creation and annihilation operators, respectively, for the $f$ orbital $u$ and spin $\sigma$ at site $i$. Specifically, we take $\tilde{t}_{iu,i'v,\sigma}$ in the form 
\begin{equation}
\tilde{t}_{iu,i'v,\sigma}=t_{iu,i'v,\sigma}+\sum_{o,p} \frac{t_{iu,op,\sigma}t^*_{i'v,op,\sigma}}{\Delta_{p\textrm{-}uv}},
\label{eq:tilde-t}
\end{equation}
where the first term $t_{iu,i'v,\sigma}$ describes the direct hopping between orbital $u$ at site $i$ and orbital $v$ at site $i'$ for spin $\sigma$, and the second term describes the indirect hoppings via oxygen $2p$ orbitals; $t_{iu,op,\sigma}$ is the transfer integral for spin $\sigma$ between $4f$ orbital $u$ at site $i$ and $2p$ orbital $p$($=x$, $y$, and $z$) at one of two ligand sites $o$($=1$ and $2$) shared by two PrO$_6$ octahedra for the sites $i$ and $i'$, and $\Delta_{p\textrm{-}uv}$ is the harmonic mean of the energies of orbitals $u$ and $v$ measured from that of $p$. 

We estimate the values of $t_{iu,i'v,\sigma}$, $t_{iu,op,\sigma}$, and $\Delta_{p\textrm{-}uv}$ by the MLWF analyses for the electronic band structure obtained by the {\it ab initio} calculations. Note that we take into account the electron hopping only between nearest-neighbor Pr pairs for simplicity. The validity of this approximation will be examined by comparing the tight-binding band structures and those obtained by the {\it ab initio} calculations (see Figs.~\ref{fig:f3} and \ref{fig:f10}). For the quasi-2D honeycomb cases, we average the values over three types of bonds by assuming $C_3$ symmetry in each honeycomb layer for simplicity, as the deviations are very small in each Pr layer (see Sec.~\ref{sec:structureoptimization}). 

\subsection{Kramers doublet}
\label{subsec:KramersDoublet}

\begin{figure}[t!]
\includegraphics[width=0.95\columnwidth]{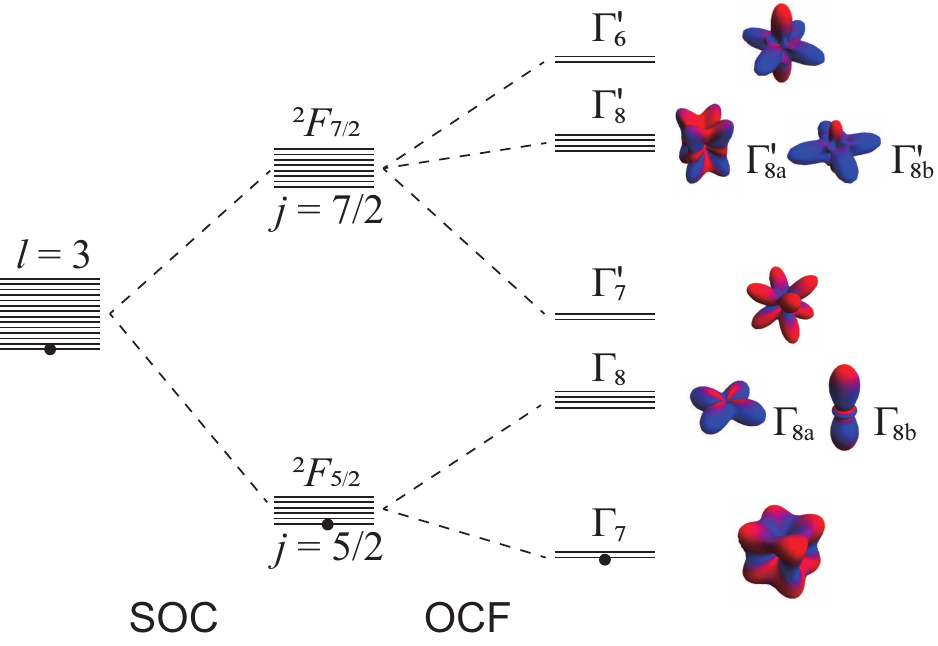}
\caption{\label{fig:f1}Energy level scheme for the $f$-orbital manifold under the spin-orbit coupling (SOC) and the octahedral crystal field (OCF). The black dot represents the occupied state in the $f^1$ electron configuration. The schematic pictures of the corresponding wave functions are also shown, where red and blue represent spin-up and spin-down density profiles, respectively.}
\end{figure} 

We consider the $4f^1$ electronic state for the mulitorbital Hubbard model in Eq.~(\ref{eq:H_multi}), namely, one $f$ electron per site on average. This is expected from the formal valence of Pr$^{4+}$ in $A_2$PrO$_3$, and indeed confirmed by the {\it ab initio} calculations in the later sections. Let us first discuss the atomic electronic state for the first two terms in Eq.~(\ref{eq:H_multi}), $\mathpzc{H}_{\textrm{SOC}}$ and $\mathpzc{H}_{\textrm{OCF}}$. The SOC in Eq.~(\ref{eq:H_SOC}) splits the 14-fold degenerate $f$-orbital manifold by the total angular momentum $j$ into the $^2F_{5/2}$ sextet with $j=5/2$ and the $^2F_{7/2}$ octet with $j=7/2$, as shown in Fig.~\ref{fig:f1} (the eigenvalues and eigenvectors are shown in Appendix~\ref{app:multiplets}). These manifolds are further split by $\mathpzc{H}_{\textrm{OCF}}$ in Eq.~(\ref{eq:H_multi}). The $j=5/2$ manifold is split into $\Gamma_7$ doublet and $\Gamma_8$ quartet, while the $j=7/2$ manifold is split into $\Gamma^{\prime}_7$ doublet, $\Gamma^{\prime}_8$ quartet, and $\Gamma_6$ doublet, as shown in Fig.~\ref{fig:f1}. The $\Gamma_7$ doublet from the $j=5/2$ manifold has the lowest eigenvalue of $\mathpzc{H}_{\textrm{OCF}}$ at $-240B_{40}$, which is described by
\begin{align}
\ket{j=\frac{5}{2}, \Gamma_7; \pm} &= \frac{1}{\sqrt{21}} (2{\rm i}c^{\dagger}_{\xi\mp} \mp 2c^{\dagger}_{\eta\mp} \pm 2{\rm i}c^{\dagger}_{\zeta\pm} + 3c^{\dagger}_{A\pm})\ket{0}.
\label{eq:G7}
\end{align}
Here, we use the cubic harmonic basis as in Eq.~(\ref{eq:Hhopij}) (we omit the site label $i$ for simplicity); $\ket{0}$ is the vacuum of $f$ electrons. (The eigenvalues and eigenvectors for the other multiplets are shown in Appendix~\ref{app:multiplets}.) The lowest-energy $\Gamma_{7}$ doublet in Eq.~(\ref{eq:G7}) comprises a time-reversal pair, which can be regarded as a pseudospin $\ket{\pm}$ with the effective angular momentum $j_{\rm eff}=1/2$. For the pseudospin state, we introduce the operator $\mathbf{S} = (S^x,S^y,S^z)^{\textrm{T}}$ defined by
\begin{equation}
S^\mu=-\frac{3}{5}
\begin{pmatrix} 
\mel{+}{j^\mu}{+} & \mel{+}{j^\mu}{-} \\
\mel{-}{j^\mu}{+}  & \mel{-}{j^\mu}{-}  
\end{pmatrix}
=\frac{1}{2}\sigma^\mu,
\label{eq:pseudospin}
\end{equation}
where $\mathbf{j}$ and $\boldsymbol{\sigma}$ are the total angular momentum operator and the Pauli matrix, respectively.

\subsection{Perturbation expansion}
\label{subsec:perturbation}

Next, for the $\Gamma_7$ doublet described by the pseudospin in Eq.~(\ref{eq:pseudospin}), we discuss the effect of the Coulomb interaction and the electron hopping described by the latter two terms in Eq.~(\ref{eq:H_multi}). We assume that the Coulomb interaction in $\mathpzc{H}_{\textrm{int}}$ is large enough to realize the spin-orbit Mott insulating state in the basis of the $\Gamma_7$ doublet, where the $4f$ electrons are localized at each site with one electron per site. For this situation, we derive the low-energy effective Hamiltonian by employing the perturbation expansion with respect to the electron hopping in $\mathpzc{H}_{\textrm{hop}}$. The lowest-order contribution is obtained from the second-order perturbation. The effective Hamiltonian for a pseudospin pair for nearest-neighbor sites $i$ and $i'$ on a $\mu$ bond is calculated by
\begin{widetext}
\begin{equation}
h^{\left( \mu \right)}_{ii'}=\sum_{a,b,c,d=\pm}\sum_{n}\frac{\mel{c,d}{\mathpzc{H}^{\left( \mu \right)}_{\textrm{hop},ii'}}{n}\mel{n}{\mathpzc{H}^{ \left( \mu \right) }_{\textrm{hop},ii'}}{a,b}}{E_0-E_n}\ket{c,d}\bra{a,b}.
\label{eq:h^(2)}
\end{equation}
\end{widetext}
where $\ket{a,b}$ and $\ket{c,d}$ are the initial and final two-site states with $4f^1$-$4f^1$ electron configurations described by the eigenvalues of the pseudospin in Eq.~(\ref{eq:pseudospin}) at each site, and $\ket{n}$ is the intermediate states with $4f^2$-$4f^0$ or $4f^0$-$4f^2$ electron configurations; $E_0$ is the energy for the initial and final states, while $E_n$ is for the intermediate state $\ket{n}$. In the present calculations, we classify the intermediate states with the $f^2$ electron configuration on the basis of the Russel-Saunders scheme by using the eigenstates of $\mathpzc{H}_{\textrm{int}}+\mathpzc{H}_{\textrm{SOC}}$ in the absence of  $\mathpzc{H}_{\textrm{CEF}}$. This results in the 91 multiplets, whose explicit forms are given with their energy eigenvalues in Appendix~\ref{app:multiplets2}.  

The effective Hamiltonian in Eq.~(\ref{eq:h^(2)}) can be summarized into the form of the spin Hamiltonian in terms of the pseudospins in Eq.~(\ref{eq:pseudospin}). The effective pseudospin Hamiltonian, e.g, for the $z$ bond, is given in the matrix form
\begin{equation}
\mathpzc{H}^{ \left( z \right) }_{ii'}=
\mathbf{S}_i^{\rm T}
\begin{bmatrix} 
\it{J} & \Gamma & \Gamma^{\prime} \\
\Gamma & \it{J} & \Gamma^{\prime} \\
\Gamma^{\prime}  & \Gamma^{\prime} & \it{J}+\it{K}
\end{bmatrix}
\mathbf{S}_{i'}.
\label{eq:Heff}
\end{equation}
The total Hamiltonian is given by the sum over the neighboring $\mu=x,y,z$ bonds as
\begin{equation}
\mathpzc{H}_{\rm eff} = \sum_{\mu} \sum_{{\langle i,i' \rangle}_{\mu}} \mathpzc{H}^{(\mu)}_{ii'}
, 
\end{equation}
where $\mathpzc{H}^{(x)}_{ii'}$ and $\mathpzc{H}^{(y)}_{ii'}$ are given by cyclic permutations of $\{ xyz \}$ in $\mathpzc{H}^{(z)}_{ii'}$. We note that the spin Hamiltonian in Eq.~(\ref{eq:Heff}) for the quasi-2D honeycomb cases does not include antisymmetric exchange interactions, such as the Dzyaloshinskii-Moriya interaction~\cite{DZ1958, MO1960}, since the lattice structures possess the inversion center at the middle of each Pr-Pr bond. This is not the case for the 3D hyperhoneycomb case, but it turns out that the antisymmetric exchange interactions are negligibly small as discussed in Sec.~\ref{sec:couplings2}. 

\section{Result} 
\label{sec:result}

In this section, we show the results for a series of the quasi-2D honeycomb compounds $A_2$PrO$_3$ (Sec.~\ref{sec:honeycombmagnets}) and the 3D hyperhoneycomb compound $\beta$-Na$_2$PrO$_3$ (Sec.~\ref{sec:hyperhoneycombmagnets}). For the quasi-2D cases, after presenting the optimized lattice structures in Sec.~\ref{sec:structureoptimization}, we show the electronic band structures for $A$=K, Rb, and Cs in Sec.~\ref{sec:electronicstructure}. We estimate the tight-binding parameters for the multiorbital Hubbard Hamiltonian from the MLWF analysis in Sec.~\ref{sec:TandSOC} and the exchange coupling constants in the effective pseudospin Hamiltonian in Sec.~\ref{sec:couplings}. We discuss the systematic evolution of the parameters, including the previous results for $A$=Li and Na~\cite{JA2019}. In particular, we identify relevant perturbation processes to the coupling constants $J$, $K$, and $\Gamma'$ in Sec.~\ref{sec:decomposition}. In Sec.~\ref{sec:magneticphase}, we calculate the ground-state phase diagram for the $J$-$K$-$\Gamma'$ model and map out the systematic evolution while changing $A$-site ions on the phase diagram. For the 3D case, we present the results in a parallel manner from Sec.~\ref{sec:latticestructure} to \ref{sec:couplings2}, by using the experimental structure for the {\it ab initio} calculations.

\subsection{Honeycomb magnets $A_2$PrO$_3$} 
\label{sec:honeycombmagnets}

\subsubsection{Lattice structure}
\label{sec:structureoptimization}

\begin{table}[t]
\caption{\label{tab:table1}Structural parameters of the optimized structures for $A_2$PrO$_3$ ($A$=K, Rb, and Cs) with $C2/m$ symmetry. See Fig.~\ref{fig:f2}(b) for the definitions of $a$, $b$, $c$, $\beta$, and $n$. The ratio $a/n$ becomes $3/\sqrt{2}\simeq 2.12$ in ideal octahedra with O$_h$ symmetry. $d_{\textrm{Pr-Pr}}$ and $\it{\theta}_{\textrm{Pr-O-Pr}}$ denote the average values of the Pr-Pr bond length and the Pr-O-Pr bond angle, respectively, for the neighboring Pr pairs within the same honeycomb layer.}
\begin{ruledtabular}
\begin{tabular}{cccc}
&K$_2$PrO$_3$&Rb$_2$PrO$_3$&Cs$_2$PrO$_3$\\
\hline
$a$ (\si{\angstrom})&6.1069&6.2158&6.3349\\
$b$ (\si{\angstrom})&10.535&10.705&10.921\\
$c$ (\si{\angstrom})&6.3442&6.6890&7.0903\\
$\it{\beta}$ (deg)&109.04&108.29&107.42\\
\hline
$n$ (\si{\angstrom})&2.4103&2.4516&2.3698\\
$a/n$&2.5337&2.5354&2.6732\\
\hline
$d_{\textrm{Pr-Pr}}$ (\si{\angstrom})&3.5188&3.5778&3.6471\\
$\it{\theta}_{\textrm{Pr-O-Pr}}$ (deg)&103.68&105.49&107.57\\
\end{tabular}
\end{ruledtabular}
\end{table}

\begin{figure}[t]
\includegraphics[width=0.9\columnwidth]{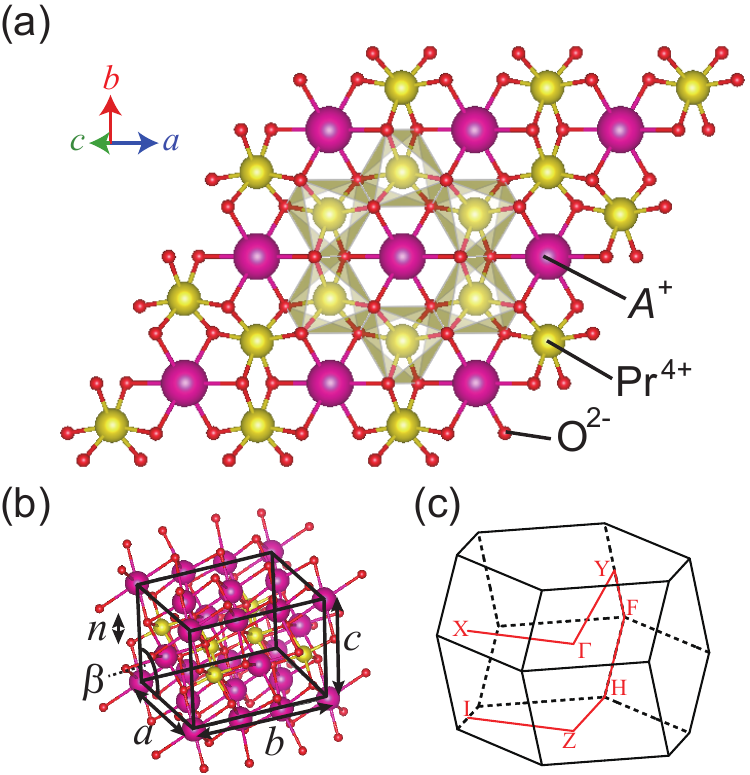}
\caption{\label{fig:f2}(a) and (b) The optimized $C2/m$ monoclinic structure for $A_2$PrO$_3$ with $A$=Rb. The other cases with $A$=K and Cs have similar structures. The purple, yellow, and red spheres denote $A^{+}$, Pr$^{4+}$, and O$^{2-}$ ions, respectively. The edge-sharing network of PrO$_6$ octahedra is partially shown. In (b), the black lines represent a primitive unit cell with the lattice parameters; $n$ is the average distance of the O layers sandwiching the Pr layer. (c) The first Brillouin zone for the monoclinic structure. The red lines represent the symmetric lines used in Fig.~\ref{fig:f3}.}
\end{figure} 

\begin{figure*}[ht!]
\includegraphics[width=2.0\columnwidth]{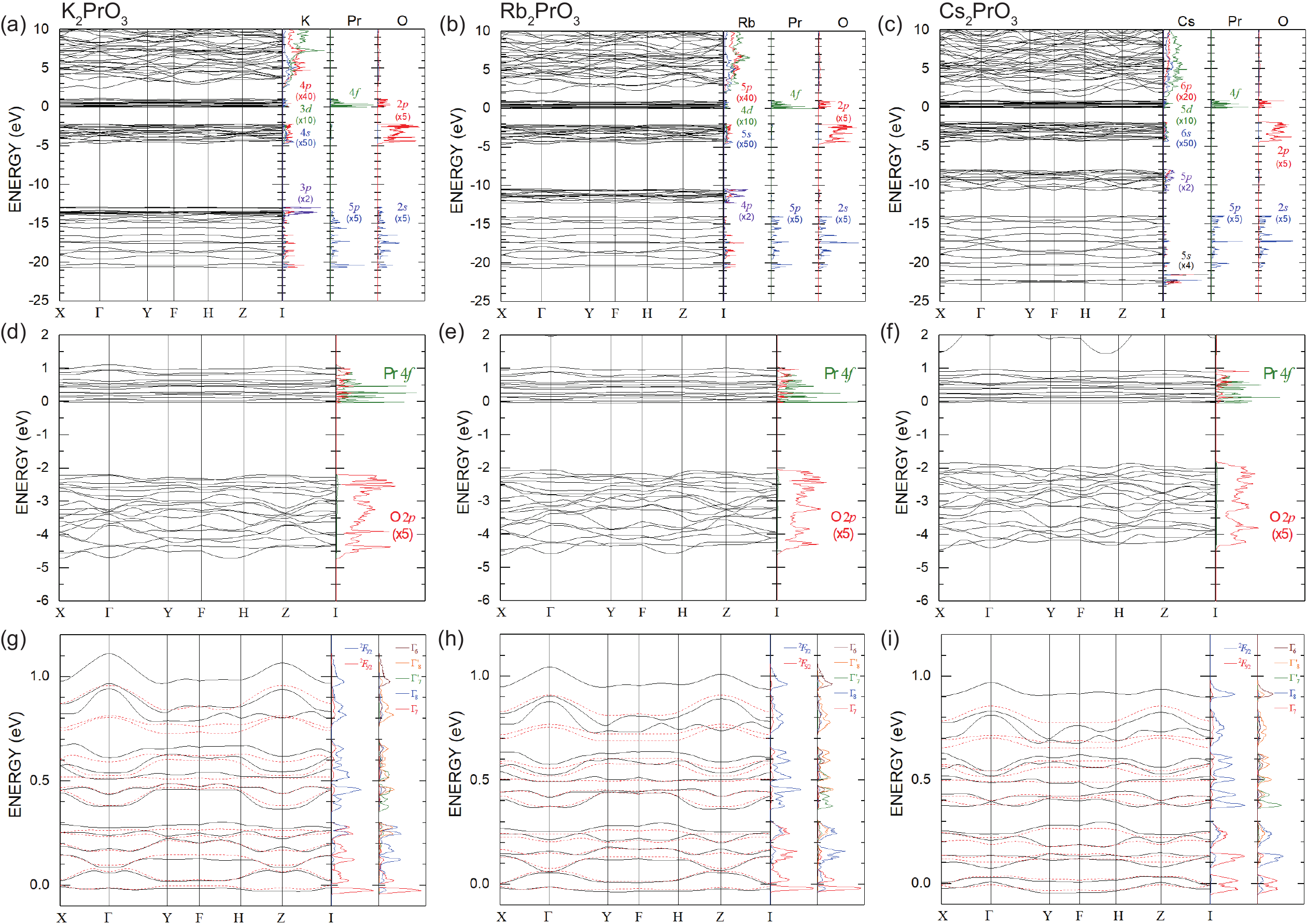}
\caption{\label{fig:f3}Electronic band structures for $A_2$PrO$_3$ obtained by the relativistic {\it ab initio} calculations: (a)(d)(g) for $A$=K, (b)(e)(h) for $A$=Rb, and (c)(f)(i) for $A$=Cs. The figures (a)-(c) are in the energy range from $-25$~eV to $10$~eV, (d)-(f) are from $-6$~eV to $2$~eV, and (g)-(i) are from $-0.1$~eV to $1.2$~eV. The band structures are drawn along the symmetric lines indicated in Fig.~\ref{fig:f2}(c). The red dashed lines in (g)-(i) show the band dispersions obtained by the tight-binding calculations with nearest-neighbor transfers estimated by the MLWFs. The right panels in each figure display the projected density of states to various orbitals of three atoms $A$, Pr, and O in (a)-(c), Pr $4f$ and O $2p$ orbitals in (d)-(f), and the $^2F_{5/2}$, $^2F_{7/2}$, $\Gamma_7$, $\Gamma_8$, $\Gamma_7'$, $\Gamma_8'$, and $\Gamma_6$ manifolds of the Pr $4f$ states in (g)-(i). The Fermi level is set to zero.}
\end{figure*} 

Table~\ref{tab:table1} summarizes the structural parameters for $A_2$PrO$_3$ ($A$=K, Rb, and Cs) with $C2/m$ symmetry obtained by the structural optimization described in Sec.~\ref{subsec:ab_initio}. The optimized structures are composed of 2D honeycomb layers with edge-sharing PrO$_6$ octahedra, as exemplified in Fig.~\ref{fig:f2} for $A$=Rb. While the $A$-site ionic radius increases, not only the intralayer Pr-Pr bond length $d_{\textrm{Pr-Pr}}$ but also the interlayer distance is elongated, as shown in Table~\ref{tab:table1}. At the same time, the value of $a/n$, which is a measure of the degree of trigonal distortions, and the Pr-O-Pr bond angle $\theta_{\textrm{Pr-O-Pr}}$ gradually deviate from the values for the ideal octahedra, $3/\sqrt{2}$ and $90$\si{\degree}, respectively~\cite{[We note that the Li and Na cases slightly deviate from the systematic change{\rm ,} presumably due to the use of a different type of the pseudopotential for the alkali metals~\cite{JA2019}. See also the footnote in Ref.~\onlinecite{DU0000}] DU0001}. Although the lattice symmetry is $C2/m$, the deviations from the perfect honeycomb structure with $C_3$ symmetry are very small in each Pr layer for all the compounds; the differences of $d_{\textrm{Pr-Pr}}$ and $\theta_{\textrm{Pr-O-Pr}}$ among the different bond directions are within $\simeq0.05$~\si{\angstrom} and $\simeq1$~\si{\degree}, respectively. 

We note that the bond lengths and angles for the cases with $A$=K, Rb, and Cs are comparatively larger than those for the $d^5$ Kitaev honeycomb magnets, $A_2$IrO$_3$ ($A$=Li and Na)~\cite{CH2012, GR2013} and $\alpha$-RuCl$_3$~\cite{JO2015}. On the other hand, Li$_2$PrO$_3$ and Na$_2$PrO$_3$, which were studied previously~\cite{JA2019}, have similar structural parameters to the $d^5$ candidates; the bond lengths for the Li and Na cases are close to those for Na$_2$IrO$_3$ and $\alpha$-RuCl$_3$, respectively, and the bond angles are close to those for $\alpha$-RuCl$_3$~\cite{JO2015} and Na$_2$IrO$_3$~\cite{CH2012, GR2013}, respectively.

\subsubsection{Electronic structure}
\label{sec:electronicstructure}

The electronic band structures and the projected density of states for nonmagnetic states of $A_2$PrO$_3$ ($A$=K, Rb, and Cs) are shown in Fig.~\ref{fig:f3}. In all the cases, the Pr $4f$ bands are well isolated from the other bands and located around the Fermi level set to zero. In the higher-energy region, there are hybridized bands of $s$, $p$, and $d$ orbitals of the $A$ cations above $2.5$~eV, $2.0$~eV, and $1.8$~eV for $A$=K, Rb, and Cs, respectively. The Pr $4s$ bands are located above 10~eV for all the compounds (not shown). Meanwhile, in the lower-energy region, the O $2p$ bands are located in the range from $-5.5$ to $-2.2$~eV for $A$=K, from $-5.2$ to $-2.0$~eV for $A$=Rb, and from $-4.5$ to $-1.8$~eV for $A$=Cs, respectively, with weak hybridization with the Pr $4f$ bands. The bands in the deeper energy range from $-21$ to $-13$~eV for $A$=K are mainly from the hybridization of K $3p$, Pr $5p$, and O $2s$ orbitals. The bands in the range from $-12.2$ to $-10.4$~eV for $A$=Rb and in the range from $-10.9$ to $-8.0$~eV for $A$=Cs are mainly from Rb $4p$ and Cs $5p$ orbitals, respectively. The bands in the range from $-21$ to $-14$~eV for $A$=Rb and Cs are mainly from the hybridization of Pr $5p$ and O $2s$ orbitals. 

Reflecting the localized nature of the $f$ orbitals, the bandwidths of the well-isolated Pr $4f$ bands are narrow. The bandwidth decreases with the increase of the $A$-site ionic radii: $\simeq 1.3$~eV for $A$=K, $\simeq 1.2$~eV for $A$=Rb, and $\simeq 1.1$~eV for $A$=Cs. This is in accordance with the increased lattice constants in Table~\ref{tab:table1}. As shown in the enlarged figures in Figs.~\ref{fig:f3}(g), \ref{fig:f3}(h), and \ref{fig:f3}(i), the $4f$ bands are split into the bands predominantly originating from the $^2F_{5/2}$ sextet (below $0.3$~eV) and those from the $^2F_{7/2}$ octet (above $0.3$~eV), as expected from the atomic level scheme under the strong SOC in Fig.~\ref{fig:f1}. These two bunches of the bands are further split under the crystal field; the $^2F_{5/2}$ bands are split into the bands dominated by the $\Gamma_7$ doublet and the $\Gamma_8$ quartet, while the $^2F_{7/2}$ bands are split into those dominated by $\Gamma_7'$ doublet, $\Gamma_8'$ quartet, and $\Gamma_6'$ doublet, as expected in Fig.~\ref{fig:f1}.

In the $4f^1$ state, the lowest-energy shallow band (doubly degenerate) below the Fermi level, which predominantly originates from the $\Gamma_7$ doublet split from the $^2F_{5/2}$ sextet, is occupied. In particular, in the $A$=K and Rb cases, the band is fully occupied, indicating that the system is a band insulator. The band gap is estimated as $\simeq 18$~meV and $9$~meV for $A$=K and Rb, respectively, Meanwhile, for the $A$=Cs case, the (second) lowest-energy band is slightly hole (electron) doped, indicating that the system is a compensated metal. Nonetheless, it is expected for all the cases that the Coulomb interactions can make the system a spin-orbit Mott insulator. 

In Figs.~\ref{fig:f3}(g), \ref{fig:f3}(h), and \ref{fig:f3}(i), we also show the tight-binding band structures with the transfer integrals between nearest-neighbor Pr cations estimated from the MLWF analysis (see the next section). The {\it ab initio} results for the $4f$ bands are well reproduced, especially for the relevant low-energy bands near the Fermi level. This indicates that further-neighbor transfer integrals are less significant, presumably due to the localized nature of the $4f$ orbitals. Based on this observation, in Sec.~\ref{sec:couplings}, we construct effective models for the $\Gamma_7$ pseudospins in Eq.~(\ref{eq:G7}) by taking into account only the nearest-neighbor transfer integrals in the same honeycomb layer.

\subsubsection{Transfer integrals and SOC}
\label{sec:TandSOC}

\begin{figure}[ht!]
\includegraphics[width=0.95\columnwidth]{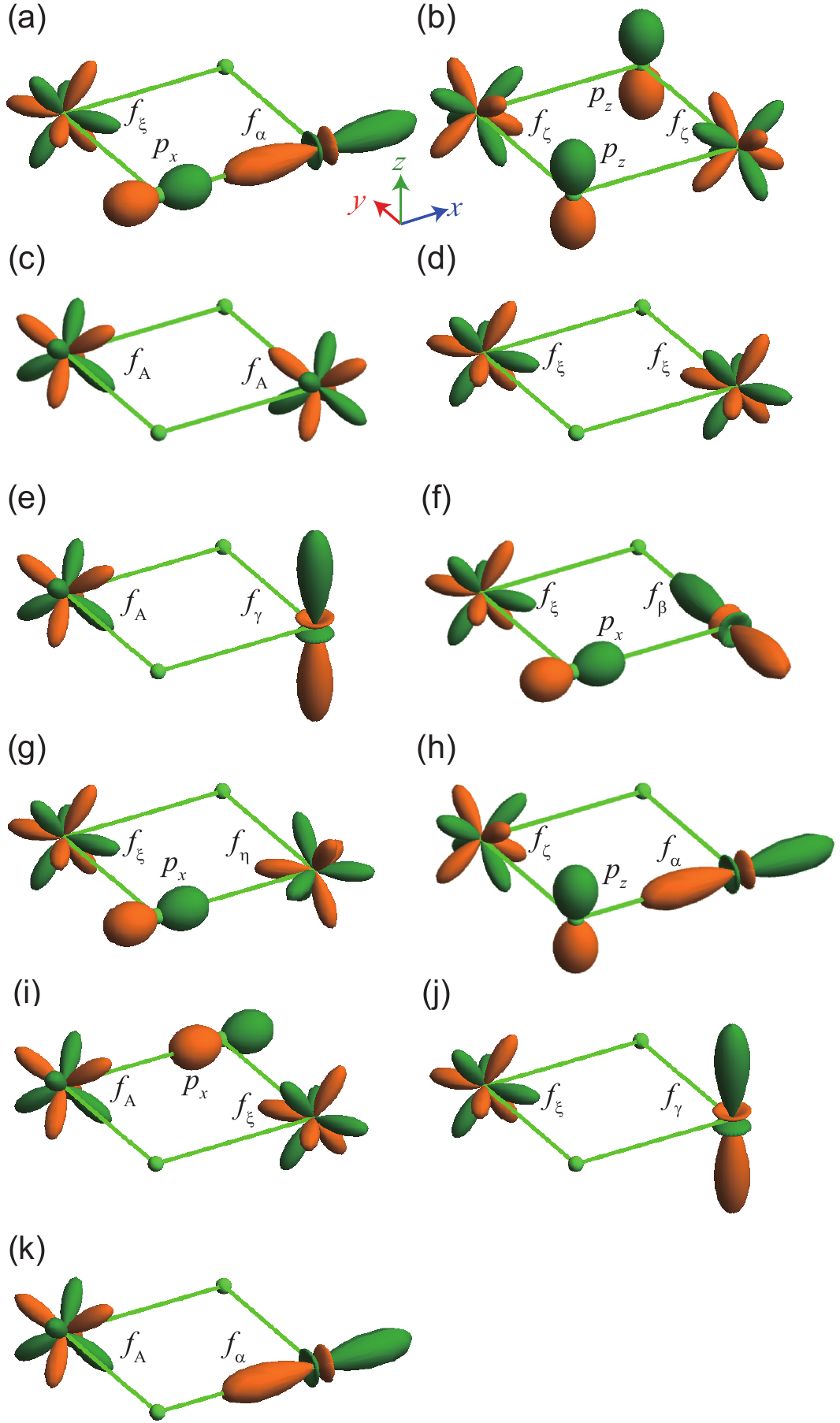}
\caption{\label{fig:f4}Relevant hopping processes along a $z$ bond: (a) indirect $f_\xi$-$p_x$-$f_\alpha$, (b) indirect $f_\zeta$-$p_z$-$f_\zeta$, (c) direct $f_A$-$f_A$, (d) direct $f_\xi$-$f_\xi$, (e) direct $f_A$-$f_\gamma$, (f) direct $f_\xi$-$f_\beta$ and indirect $f_\xi$-$p_x$-$f_\beta$, (g) direct $f_\xi$-$f_\eta$ and indirect $f_\xi$-$p_x$-$f_\eta$, (h) indirect $f_\zeta$-$p_z$-$f_\alpha$, (i) indirect $f_A$-$p_x$-$f_\xi$, (j) direct $f_\xi$-$f_\gamma$, and (k) direct $f_A$-$f_\alpha$.} 
\end{figure}

\begin{widetext}
\begin{table*}[t]
\centering
\caption{\label{tab:table2}Nearest-neighbor transfer integrals $\tilde{t}_{iu,i'v,+}$ on a $z$ bond for $A_2$PrO$_3$ ($A$=K, Rb, and Cs); $u$ is in the row and $v$ is in the column. $\tilde{t}_{iu,i'v,-}$ are given by the complex conjugates. The unit is in meV. The upper-right half of the table is omitted as the matrix is Hermite conjugate.}
\begin{ruledtabular}
\begin{tabular}{cccccccc}
$A$=K&$\xi$&$\eta$&$\zeta$&$A$&$\alpha$&$\beta$&$\gamma$\\
\hline
$\xi$&$12.3$& & & & & & \\
$\eta$&$-7.16+0.15{\rm i}$&$12.3$& & & & & \\
$\zeta$&$-1.04+0.12{\rm i}$&$-1.04+0.12{\rm i}$&$-84.1$& & & & \\
$A$&$5.30-0.09{\rm i}$&$-5.30-0.09{\rm i}$&$0.02+0.68{\rm i}$&$-30.0$& & & \\
$\alpha$&$-55.9+0.17{\rm i}$&$12.4+0.25{\rm i}$&$20.5-0.27{\rm i}$&$-10.6+0.22{\rm i}$&$123$& & \\
$\beta$&$-12.4+0.25{\rm i}$&$55.9+0.17{\rm i}$&$-20.5-0.27{\rm i}$&$-10.6-0.22{\rm i}$&$-49.1- 0.88{\rm i}$&$123$& \\
$\gamma$&$3.21+0.49{\rm i}$&$-3.21+0.49{\rm i}$&$-0.01-0.17{\rm i}$&$-7.32$&$7.53+0.53{\rm i}$&$7.53-0.53{\rm i}$&$44.2$\\ 
\hline \hline
$A$=Rb&$\xi$&$\eta$&$\zeta$&$A$&$\alpha$&$\beta$&$\gamma$\\
\hline
$\xi$&$9.15$& & & & & & \\
$\eta$&$-11.5+0.09{\rm i}$&$9.15$& & & & & \\
$\zeta$&$-0.89-0.21{\rm i}$&$-0.89+0.21{\rm i}$&$-79.4$& & & & \\
$A$&$8.48-0.03{\rm i}$&$-8.48-0.03{\rm i}$&$-0.63{\rm i}$&$-22.5$& & & \\
$\alpha$&$-47.1+0.06{\rm i}$&$12.1+0.24{\rm i}$&$35.0-0.34{\rm i}$&$-14.3+0.14{\rm i}$&$110$& & \\
$\beta$&$-12.1+0.24{\rm i}$&$47.1+0.06{\rm i}$&$-35.0-0.34{\rm i}$&$-14.3-0.14{\rm i}$&$-46.7- 0.68{\rm i}$&$110$& \\
$\gamma$&$5.98+0.49{\rm i}$&$-5.98+0.49{\rm i}$&$0.02{\rm i}$&$-1.25$&$11.2+0.55{\rm i}$&$11.2-0.55{\rm i}$&$40.3$\\ 
\hline \hline
$A$=Cs&$\xi$&$\eta$&$\zeta$&$A$&$\alpha$&$\beta$&$\gamma$\\
\hline
$\xi$&$5.22$& & & & & & \\
$\eta$&$-16.04+0.04{\rm i}$&$5.22$& & & & & \\
$\zeta$&$-0.09-0.34{\rm i}$&$-0.09+0.34{\rm i}$&$-68.2$& & & & \\
$A$&$12.1+0.05{\rm i}$&$-12.1+0.05{\rm i}$&$-0.54{\rm i}$&$-12.5$& & & \\
$\alpha$&$-37.3+0.05{\rm i}$&$9.46+0.26{\rm i}$&$50.6-0.36{\rm i}$&$-16.5+0.10{\rm i}$&$91.6$& & \\
$\beta$&$-9.46+0.26{\rm i}$&$37.3+0.05{\rm i}$&$-50.6-0.36{\rm i}$&$-16.5-0.10{\rm i}$&$-35.4- 0.42{\rm i}$&$91.6$& \\
$\gamma$&$7.80+0.45{\rm i}$&$-7.80+0.45{\rm i}$&$0.23{\rm i}$&$5.20$&$16.4+0.48{\rm i}$&$16.4-0.48{\rm i}$&$33.1$\\ 
\end{tabular}
\end{ruledtabular}
\end{table*}
\end{widetext}

Performing the MLWF analyses on the {\it ab initio} band structures, we estimate the transfer integrals between the Pr cations. The results for nearest-neighbor pairs on a $z$ bond are presented in Table~\ref{tab:table2}. Among the matrix elements, 11 types give relevant contributions to the effective pseudospin Hamiltonian derived in Sec.~\ref{sec:couplings}: 
$\tilde{t}_{i\xi,i'\alpha,\sigma}=-\tilde{t}^*_{i\eta,i'\beta,\sigma}$, 
$\tilde{t}_{i\zeta,i'\zeta,\sigma}$, 
$\tilde{t}_{iA,i'A,\sigma}$, 
$\tilde{t}_{i\xi,i'\xi,\sigma}=\tilde{t}^*_{i\eta,i'\eta,\sigma}$, 
$\tilde{t}_{iA,i'\gamma,\sigma}$,
$\tilde{t}_{i\xi,i'\beta,\sigma}=-\tilde{t}^*_{i\eta,i'\alpha,\sigma}$, 
$\tilde{t}_{i\xi,i'\eta,\sigma}$, 
$\tilde{t}_{i\zeta,i'\alpha,\sigma}=-\tilde{t}^*_{i\zeta,i'\beta,\sigma}$, 
$\tilde{t}_{iA,i'\xi,\sigma}=-\tilde{t}^*_{iA,i'\eta,\sigma}$, 
$\tilde{t}_{i\xi,i'\gamma,\sigma}=-\tilde{t}^*_{i\eta,i'\gamma,\sigma}$, and
$\tilde{t}_{iA,i'\alpha,\sigma}=\tilde{t}^*_{iA,i'\beta,\sigma}$. 
Note that the transfer integrals between the $T_{1u}$ orbitals, $f_{\alpha}$, $f_{\beta}$, and $f_{\gamma}$, some of which have large amplitudes, are irrelevant since they are not involved in the $\Gamma_7$ state in Eq.~(\ref{eq:G7}). 

Let us explain how these transfer integrals arise. Suppose that trigonal distortions are absent, indirect hopping paths via O $p$ orbitals yield nonzero values of two types of transfer integrals, $\tilde{t}_{i\xi,i'\alpha,\sigma}$ with $pf\pi$ and $pf\sigma$ bonds as exemplified for the indirect hopping process $f_\xi$-$p_x$-$f_\alpha$ in Fig.~\ref{fig:f4}(a) and $\tilde{t}_{i\zeta,i'\zeta,\sigma}$ with $pf\pi$ bonds via $p_z$ as shown in Fig.~\ref{fig:f4}(b). For these two, there are also contributions from direct hopping paths between the $f$ orbitals, dominantly with $ff\pi$ and $ff\phi$ bonds for the former and an $ff\delta$ bond for the latter. Meanwhile, the direct hopping paths yield other five nonzero transfer integrals: $\tilde{t}_{iA,i'A,\sigma}$ with an $ff\pi$ bond [Fig.~\ref{fig:f4}(c)], $\tilde{t}_{i\xi,i'\xi,\sigma}$ dominantly with $ff\sigma$ and $ff\phi$ bonds [Fig.~\ref{fig:f4}(d)], $\tilde{t}_{iA,i'\gamma,\sigma}$ with $ff\pi$ and $ff\phi$ bonds [Fig.~\ref{fig:f4}(e)], $\tilde{t}_{i\xi,i'\beta,\sigma}$ dominantly with $ff\pi$ and $ff\phi$ bonds [Fig.~\ref{fig:f4}(f)], and $\tilde{t}_{i\xi,i'\eta,\sigma}$ dominantly with $ff\sigma$ and $ff\phi$ bonds [Fig.~\ref{fig:f4}(g)]. We note that, when trigonal distortions are introduced, the indirect hopping processes $f_\xi$-$p_x$-$f_\beta$ and $f_\xi$-$p_x$-$f_\eta$ (equivalently, $f_\xi$-$p_y$-$f_\eta$) become dominant for the latter two $\tilde{t}_{i\xi,i'\beta,\sigma}$ and $\tilde{t}_{i\xi,i'\eta,\sigma}$, respectively [Figs.~\ref{fig:f4}(f) and \ref{fig:f4}(g)]. The remaining four types of the transfer integrals $\tilde{t}_{i\zeta,i'\alpha,\sigma}$, $\tilde{t}_{iA,i'\xi,\sigma}$, $\tilde{t}_{i\xi,i'\gamma,\sigma}$, and $\tilde{t}_{iA,i'\alpha,\sigma}$ become nonzero only in the presence of trigonal distortions; the indirect hopping processes $f_\zeta$-$p_z$-$f_\alpha$ and $f_\xi$-$p_x$-$f_A$ become dominant for $\tilde{t}_{i\zeta,i'\alpha,\sigma}$ and $\tilde{t}_{iA,i'\xi,\sigma}$, respectively [see Figs.~\ref{fig:f4}(h) and \ref{fig:f4}(i)], and the direct hopping processes $f_\xi$-$f_\gamma$ and $f_A$-$f_\alpha$ become dominant for $\tilde{t}_{i\xi,i'\gamma,\sigma}$ and $\tilde{t}_{iA,i'\alpha,\sigma}$, respectively [see Figs.~\ref{fig:f4}(j) and \ref{fig:f4}(k)].

The amplitudes of $\tilde{t}_{i\xi,i'\alpha,\sigma}$ and $\tilde{t}_{i\zeta,i'\zeta,\sigma}$, which are dominated by the indirect hopping paths, are quite large among the 11 types of transfer integrals. They, however, decrease with the increase in the $A$-site ionic radii which enhance the trigonal distortions. On the other hand, among the five transfer integrals predominantly originating from the direct hopping paths, the amplitude of $\tilde{t}_{iA,i'A,\sigma}$ is distinctively large, which also decreases with the increase of the $A$-site ionic radii due to the increase in $d_{\textrm{Pr-Pr}}$. While the amplitudes of the remaining four $\tilde{t}_{i\zeta,i'\alpha,\sigma}$, $\tilde{t}_{iA,i'\xi,\sigma}$, $\tilde{t}_{i\xi,i'\gamma,\sigma}$, and $\tilde{t}_{iA,i'\alpha,\sigma}$ become larger with the increase of trigonal distortions for larger $A$-site ionic radii, that of $\tilde{t}_{i\zeta,i'\alpha,\sigma}$ via the $pf\sigma$ bond between $p_z$ and $f_\alpha$ orbitals is particularly sensitive and becomes largest for $A$=Cs. 

In addition to the transfer integrals, we estimate the SOC coefficient $\lambda$ in Eq.~(\ref{eq:H_SOC}) from the comparison of the band structures in Fig.~\ref{fig:f3} with those obtained by non-relativistic calculations. The values of $\lambda$ are estimated as $\lambda \simeq0.12$~eV for $A$=K (same for $A$=Li and Na~\cite{JA2019}) and $\lambda \simeq0.11$~eV for $A$=Rb and Cs. We note that these are close to the empirical values~\cite{HI1994, PO1996}. 

\subsubsection{Effective exchange couplings}
\label{sec:couplings}

\begin{figure}[h!]
\includegraphics[width=0.85\columnwidth]{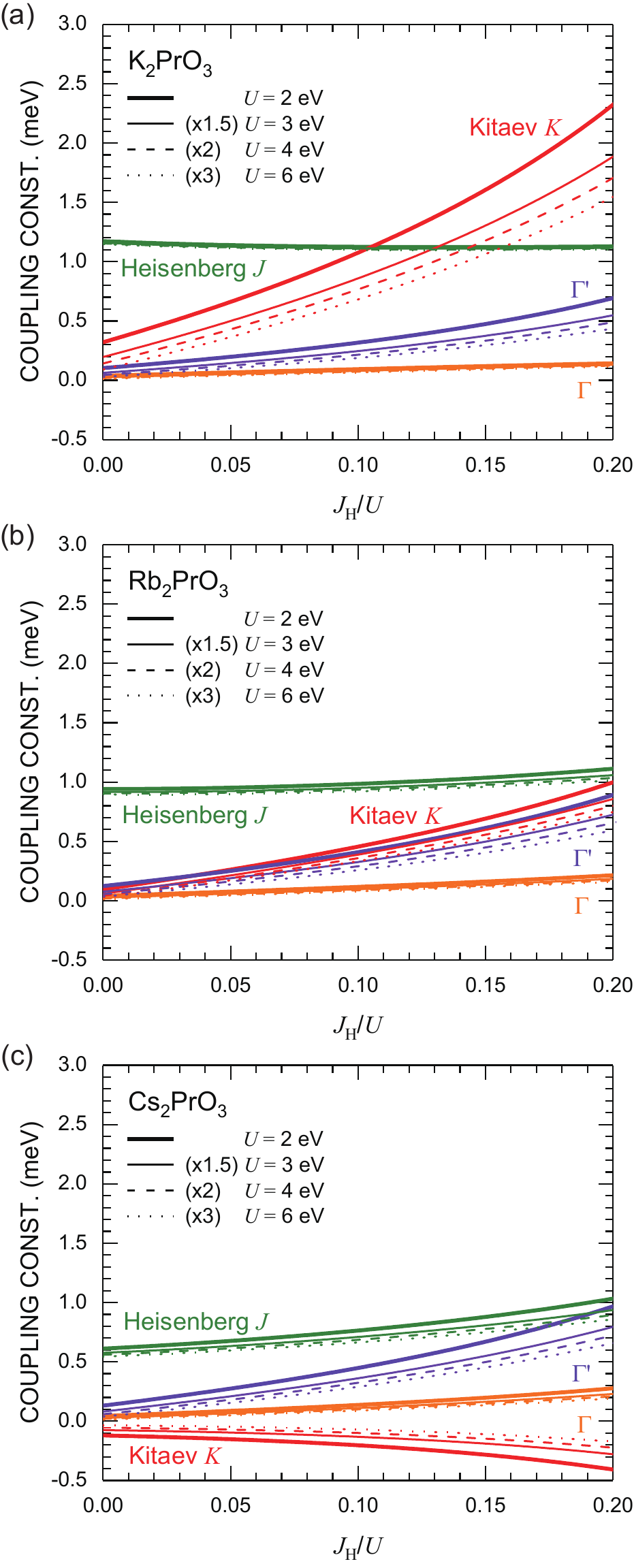}
\caption{\label{fig:f5}Coupling constants in the effective pseudospin Hamiltonian in Eq.~(\ref{eq:Heff}) for (a) K$_2$PrO$_3$, (b) Rb$_2$PrO$_3$, and (c) Cs$_2$PrO$_3$ as functions of the Hund's-rule coupling $J_{\rm H}$. The green, red, orange, and purple lines represent the Heisenberg $J$, Kitaev $K$, and off-diagonal couplings $\Gamma$ and $\Gamma^{\prime}$, respectively. The bold, solid, dashed, and dotted lines represent the results at $U=2$, $3$, $4$, and $6$~eV, respectively. The SOC coefficient $\lambda$ is set to $0.12$~eV for K$_2$PrO$_3$ and $0.11$~eV for Rb$_2$PrO$_3$ and Cs$_2$PrO$_3$, respectively.}
\end{figure} 

\begin{figure*}[th!]
\includegraphics[width=1.8\columnwidth]{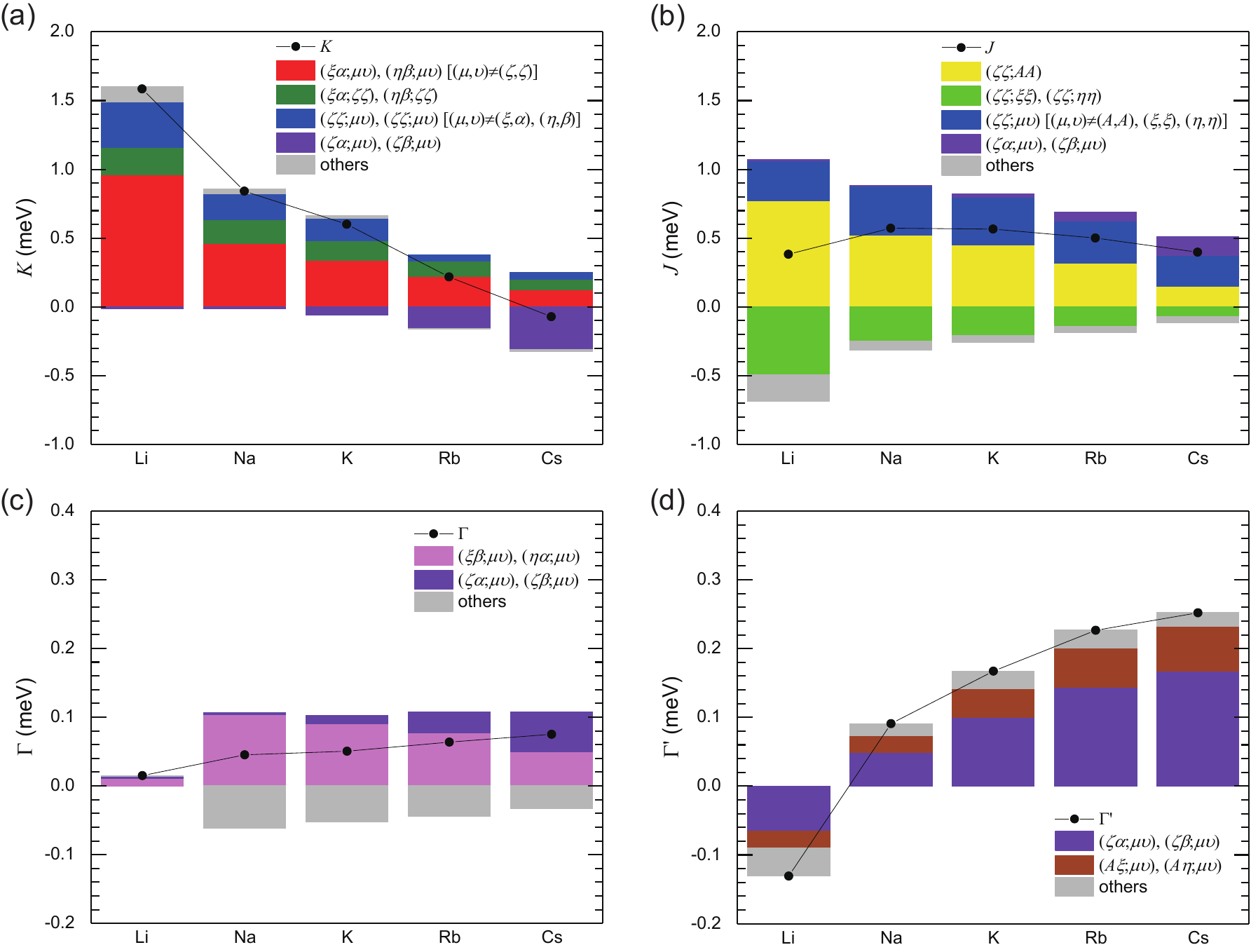}
\caption{\label{fig:f6}Coupling constants decomposed into the contributions from different hopping processes in the perturbation; ($\tilde{\mu}\tilde{\nu}$;$\mu\nu$) represents the contribution via two hopping integrals $\tilde{t}_{i\tilde{\mu},i'\tilde{\nu},\sigma}$ and $\tilde{t}_{i\mu,i'\nu,\sigma}$ on a $z$ bond ($\tilde{\mu}, \tilde{\nu}, \mu, \nu = \xi, \eta, \zeta, A, \alpha, \beta, \gamma$). The results are shown for $A_2$PrO$_3$ ($A$=Li, Na, K, Rb, and Cs): (a) Kitaev $K$, (b) Heisenberg $J$, and symmetric off-diagonal couplings (c) $\Gamma$ and (d) $\Gamma^{\prime}$. See also Ref.~\onlinecite{JA2019} for the data for $A$=Li and Na. The black points indicate the net values of the coupling constants. The onsite Coulomb energy $U$ and the Hund's-rule coupling $J_{\rm H}/U$ are set to $4$~eV and $0.15$, respectively, for all compounds. The SOC coefficient $\lambda$ is set to $0.12$~eV for $A$=Li, Na, and K, and $0.11$~eV for $A$=Rb and Cs.} 
\end{figure*}

\begin{figure*}
\includegraphics[width=1.9\columnwidth]{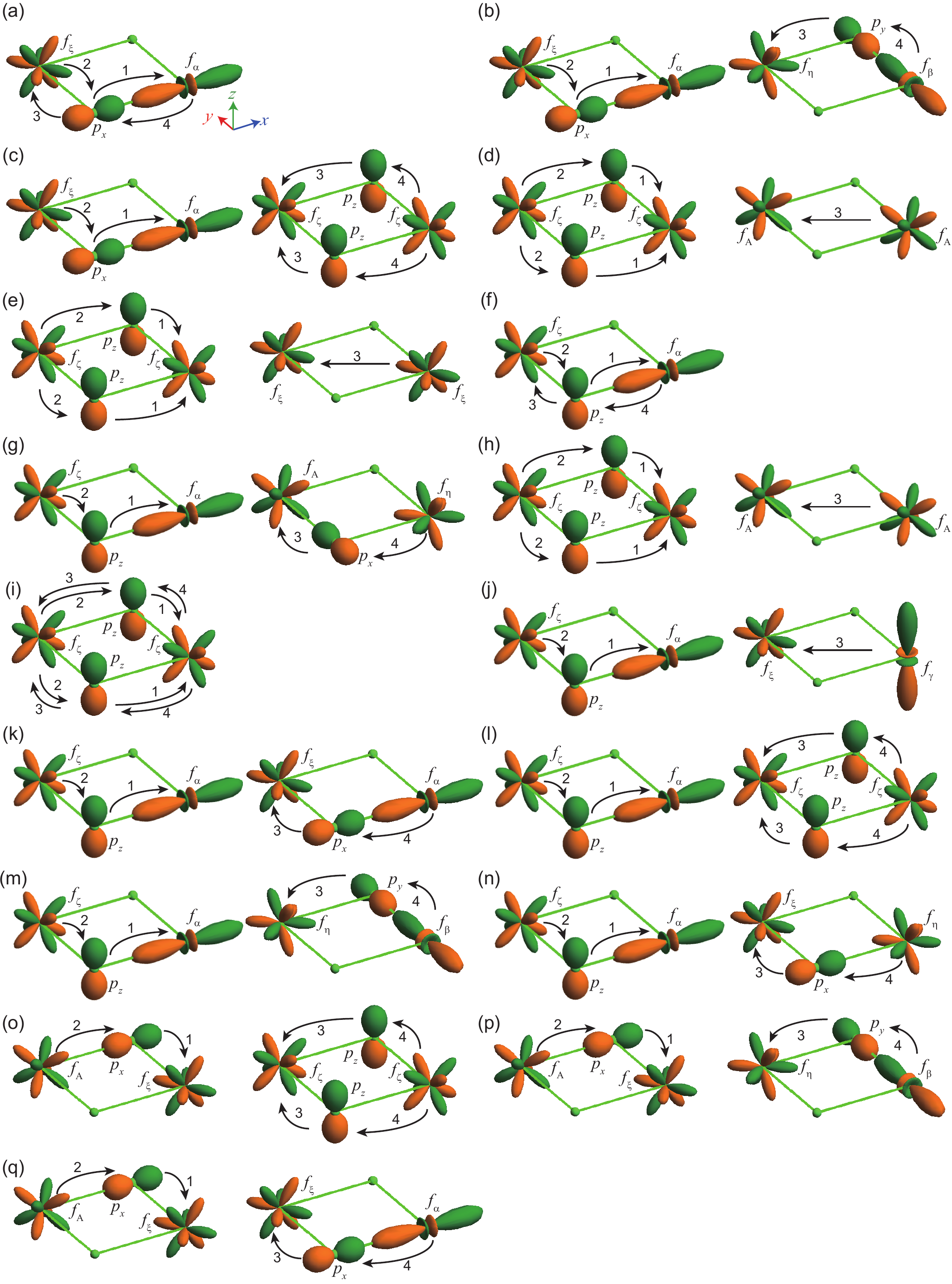}
\caption{\label{fig:f7}Examples of ($\tilde{\mu}\tilde{\nu}$;$\mu\nu$) that give dominant contributions to the effective coupling constants: (a) ($\xi\alpha$;$\xi\alpha$), (b) ($\xi\alpha$;$\eta\beta$), (c) ($\xi\alpha$;$\zeta\zeta$), and (d) ($\zeta\zeta$;$AA$) contributing to the AMF $K$, (e) ($\zeta\zeta$;$\xi\xi$) to the AFM $K$ and the FM $J$, (f) ($\zeta\alpha$;$\zeta\alpha$) to the FM $K$ and the AFM $J$, (g) ($\zeta\alpha$;$A\eta$) to the FM $K$ and the AFM $J$, (h) ($\zeta\zeta$;$AA$), (i) ($\zeta\zeta$;$\zeta\zeta$), and (j) ($\zeta\alpha$;$\xi\gamma$) to the AFM $J$, and (k) ($\zeta\alpha$;$\xi\alpha$), (l) ($\zeta\alpha$;$\zeta\zeta$), (m) ($\zeta\alpha$;$\eta\beta$), (n) ($\zeta\alpha$;$\xi\eta$), (o) ($A\xi$;$\zeta\zeta$), (p) ($A\xi$;$\eta\beta$), and (q) ($A\xi$;$\xi\alpha$) to the positive $\Gamma'$.}
\end{figure*} 

Following the procedure of the perturbation expansion in Sec.~\ref{subsec:perturbation}, we estimate the coupling constants in the effective pseudospin Hamiltonian in Eq.~(\ref{eq:Heff}). The results for $A_2$PrO$_3$ ($A$=K, Rb, and Cs) are plotted in Fig.~\ref{fig:f5} for several $U$ as functions of the ratio of the Hund's-rule coupling $J_{\rm H}$ to the onsite Coulomb repulsion $U$. The value of the effective $U$ in the transition from $4f^1$-$4f^1$ to $4f^2$-$4f^0$ was experimentally determined as $3.7$ -- $5.4$~eV with the x-ray photoelectron spectroscopy~\cite{LA1981}, while the effective $J_{\rm H}$ for $4f^2$ has been measured as $0.6$ -- $0.9$~eV with spectroscopic methods for Pr$^{3+}$ ions~\cite{CA1968, LA1982, CA1989, OG1991}. On the other hand, theoretical estimates were given as $U=5.0$ -- $5.7$~eV by the thermodynamic approximation method and the relativistic Hartree-Fock method~\cite{JO1979, HE1987} and $J_{\rm H}=0.6$ -- $1.1$~eV by the hydrogenic method and the relativistic Hartree-Fock method~\cite{JU1961, FE1962}. We note that the previous {\it ab initio} studies on Ce- and Pr-based materials with $4f^1$ or $4f^2$ electron configurations have been implemented with $U=2.5$ -- $8$~eV and $J_{\rm H}/U=0.08$ -- $0.20$~\cite{DI2005, LA2007, TR2008, YI2008, JI2009, ZH2009, NI2013, LA2015, HU2016, LO2016}. Considering these experimental and theoretical estimates, we take the range of $U=2$ -- $6$~eV and $J_{\rm H}/U=0.0$ -- $0.2$, in addition to $\lambda=0.12$~eV for $A$=K and $\lambda=0.11$~eV for $A$=Rb and Cs which are obtained in the MLWF analyses in the previous section.

In the case of K$_2$PrO$_3$ in Fig.~\ref{fig:f5}(a), the AFM Kitaev coupling $K$ is most dominant for large $J_{\rm H}/U$, while the AFM $K$ becomes smaller than the AFM Heisenberg coupling $J$ for Rb$_2$PrO$_3$ in Fig.~\ref{fig:f5}(b), and even negative (FM) for Cs$_2$PrO$_3$ in Fig.~\ref{fig:f5}(c). Combining with the previous results for Li$_2$PrO$_3$ and Na$_2$PrO$_3$~\cite{JA2019}, we find that the AFM Kitaev coupling $K$ is reduced systematically with the increase of the $A$-site ionic radii. On the other hand, the AFM Heisenberg coupling $J$ shows smaller changes and remains most relevant in the case of $A$=Rb and Cs in the entire range of $U$ and $J_{\rm H}$ studied here. The symmetric off-diagonal coupling $\Gamma^{\prime}$ is always positive for $A$=K, Rb, and Cs and gives subdominant contributions for larger $A$-site ionic radii, while $\Gamma$ is smallest in all the cases. 

Thus, we conclude that the effective pseudospin Hamiltonian for $A_2$PrO$_3$ ($A$=Li, Na, K, Rb, and Cs) can be well described by the three dominant exchange couplings $J$, $K$, and $\Gamma'$. The Kitaev coupling $K$ is AFM, except for the $A$=Cs case. The situation is in stark contrast to the $d^5$ cases where the dominant couplings are $J$, $K$, and $\Gamma$, and the Kitaev coupling $K$ is FM. The $d^5$ case was often studied by the model called the $J$-$K$-$\Gamma$ model with FM $K$~\cite{RA2014, RU2019}. Our results suggest that the present $4f^1$ case is well described by the $J$-$K$-$\Gamma'$ model with AFM $K$. In particular, as $\Gamma'$ is very small for the Li and Na cases~\cite{JA2019}, these are approximately described by the $J$-$K$ model (the Heisenberg-Kitaev model). We will show the systematic changes of the coupling constatnts on the ground-state phase diagram for the $J$-$K$-$\Gamma'$ model in Sec.~\ref{sec:magneticphase}.

\subsubsection{Decomposition into different perturbation processes}
\label{sec:decomposition}

In order to understand the origin of each coupling constant, we decompose the contributions into different perturbation processes. Figure~\ref{fig:f6} shows the decomposition into different hopping processes in the perturbation on a $z$ bond. Here, ($\tilde{\mu}\tilde{\nu}$;$\mu\nu$) denotes the contribution from the hopping process via $\tilde{t}_{i\tilde{\mu},i'\tilde{\nu},\sigma}$ and $\tilde{t}_{i\mu,i'\nu,\sigma}$ [see Eqs.~(\ref{eq:tilde-t}) and (\ref{eq:h^(2)})]. 

As shown in Fig.~\ref{fig:f6}(a), the major contributions to the AFM Kitaev coupling $K$ come from ($\xi\alpha$;$\mu\nu$) [and symmetrically equivalent ($\eta\beta$;$\mu\nu$)]. For this type, we find that the dominant contributions are from ($\xi\alpha$;$\xi\alpha$) [see Fig.~\ref{fig:f7}(a)], ($\xi\alpha$;$\eta\beta$) [Fig.~\ref{fig:f7}(b)], and ($\xi\alpha$;$\zeta\zeta$) [Fig.~\ref{fig:f7}(c)]. There are also substantial contributions from ($\zeta\zeta$;$\mu\nu$), especially ($\zeta\zeta$;$AA$) [Fig.~\ref{fig:f7}(d)] and ($\zeta\zeta$;$\xi\xi$) [Fig.~\ref{fig:f7}(e)]. For larger $A$-site ionic radii, $K$ is turned into FM mainly due to the contributions from ($\zeta\alpha$;$\mu\nu$). For this type, we find that the dominant contributions are from ($\zeta\alpha$;$\zeta\alpha$) [Fig.~\ref{fig:f7}(f)] and ($\zeta\alpha$;$A\eta$) [Fig.~\ref{fig:f7}(g)].

Figure~\ref{fig:f6}(b) displays the decomposition of the Heisenberg coupling $J$. The AFM $J$ predominantly comes from ($\zeta\zeta$;$\mu\nu$), where ($\mu$,$\nu$)$\neq$($\xi$,$\xi$) and ($\eta$,$\eta$). For this type, we find that the dominant contributions are from ($\zeta\zeta$;$AA$) [Fig.~\ref{fig:f7}(h)] and ($\zeta\zeta$;$\zeta\zeta$) [Fig.~\ref{fig:f7}(i)]. We also find that ($\zeta\alpha$;$\zeta\alpha$) [Fig.~\ref{fig:f7}(f)], ($\zeta\alpha$;$A\eta$) [Fig.~\ref{fig:f7}(g)], and ($\zeta\alpha$;$\xi\gamma$) [Fig.~\ref{fig:f7}(j)] contribute to the AFM $J$ for the compounds with large $A$-site ionic radii. We note that there are FM contributions to $J$ dominantly from ($\zeta\zeta$;$\xi\xi$) [Fig.~\ref{fig:f7}(e)] and symmetrically equivalent ($\zeta\zeta$;$\eta\eta$). 

Figures~\ref{fig:f6}(c) and \ref{fig:f6}(d) show the decompositions of the symmetric off-diagonal $\Gamma$ and $\Gamma'$, respectively. Although $\Gamma$ is always small as mentioned above, the major contributions come from the types of ($\xi\beta$;$\mu\nu$) and ($\zeta\alpha$;$\mu\nu$), where the dominant ones in the former are ($\xi\beta$;$\eta\beta$) and ($\xi\beta$;$\zeta\zeta$), and those in the latter are ($\zeta\alpha$;$A\alpha$), ($\zeta\alpha$;$\zeta\beta$), and ($\zeta\alpha$;$A\xi$). On the other hand, the major contributions to $\Gamma^{\prime}$ are mainly by ($\zeta\alpha$;$\mu\nu$) and symmetrically equivalent ($\zeta\beta;\mu\nu$). We find that the dominant contributions in ($\zeta\alpha$;$\mu\nu$) are ($\zeta\alpha$;$\xi\alpha$) [Fig.~\ref{fig:f7}(k)], ($\zeta\alpha$;$\zeta\zeta$) [Fig.~\ref{fig:f7}(l)], ($\zeta\alpha$;$\eta\beta$) [Fig.~\ref{fig:f7}(m)], and ($\zeta\alpha$;$\xi\eta$) [Fig.~\ref{fig:f7}(n)]. We also find subdominant contributions from the type ($A\xi$;$\mu\nu$), especially ($A\xi$;$\zeta\zeta$) [Fig.~\ref{fig:f7}(o)], ($A\xi$;$\eta\beta$) [Fig.~\ref{fig:f7}(p)], and ($A\xi$;$\xi\alpha$) [Fig.~\ref{fig:f7}(q)]. 

Summarizing the above analysis, we conclude that the contributions from ($\xi\alpha$;$\mu\nu$) and ($\zeta\zeta$;$\mu\nu$) play a major role in the dominant AFM Kitaev coupling $K$, while the latter ($\zeta\zeta$;$\mu\nu$) simultaneously gives a relevant contribution to the dominant AFM Heisenberg coupling $J$. We also find that, when the trigonal distortions become larger with the increase of the $A$-site ionic radii, the contribution from ($\zeta\alpha$;$\mu\nu$) becomes more relevant to all the coupling constants; in particular, it changes the sign of $K$ from AFM to FM.

\begin{figure}[t]
\includegraphics[width=0.9\columnwidth]{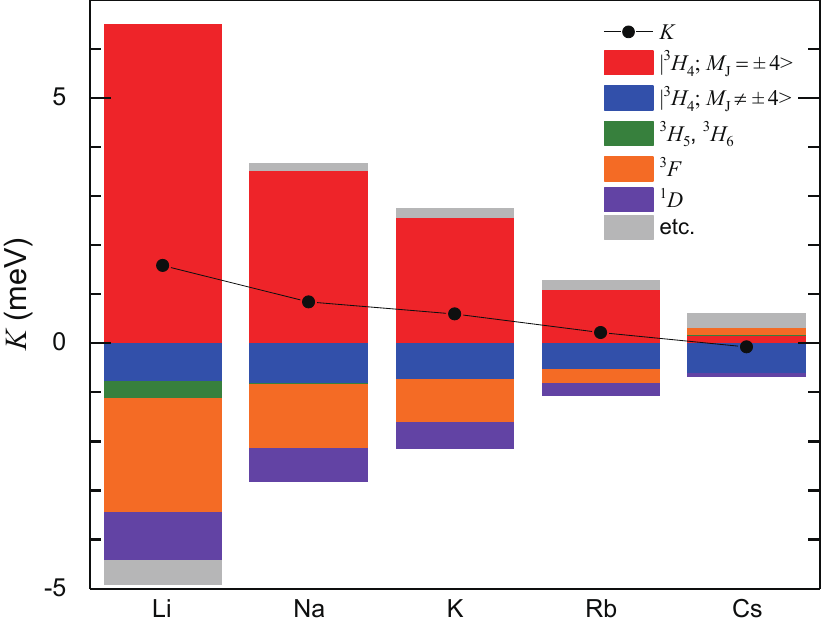}
\caption{\label{fig:f8}Coupling constants decomposed into the contributions from different intermediate $4f^2$-$4f^0$ states to the Kitaev coupling $K$ for $A_2$PrO$_3$ ($A$=Li, Na, K, Rb, and Cs). The parameters are set in the same way as in Fig.~\ref{fig:f6}.}
\end{figure} 

Finally, to further analyze the origin of the AFM $K$, we decompose $K$ into the contributions from different intermediate $4f^2$-$4f^0$ states in the perturbation [see Eq.~(\ref{eq:h^(2)})]. The result is shown in Fig.~\ref{fig:f8}. We find that the dominant contributions come from the intermediate states $^3H$ and $^3F$. These two have the lowest and second-lowest energies among many intermediate states [see Eqs.~(\ref{eq:E-3H4})-(\ref{eq:E-1S0})]. More specifically, the state $\ket{^3H_4; M_J=\pm4}$ contributes to the AFM $K$, while $\ket{^3H_4; M_J\neq\pm4}$ and $^3F$ contribute to the FM $K$. The contributions from $\ket{^3H_4; M_J=\pm4}$ and $^3F$ quickly decrease with the increase of the $A$-site ionic radii, but that from $\ket{^3H_4; M_J\neq\pm4}$ does not change so much, which finally leads to the FM $K$ in the Cs case.

\subsubsection{Possible magnetic phases}
\label{sec:magneticphase}

\begin{figure}[h!]
\includegraphics[width=0.85\columnwidth]{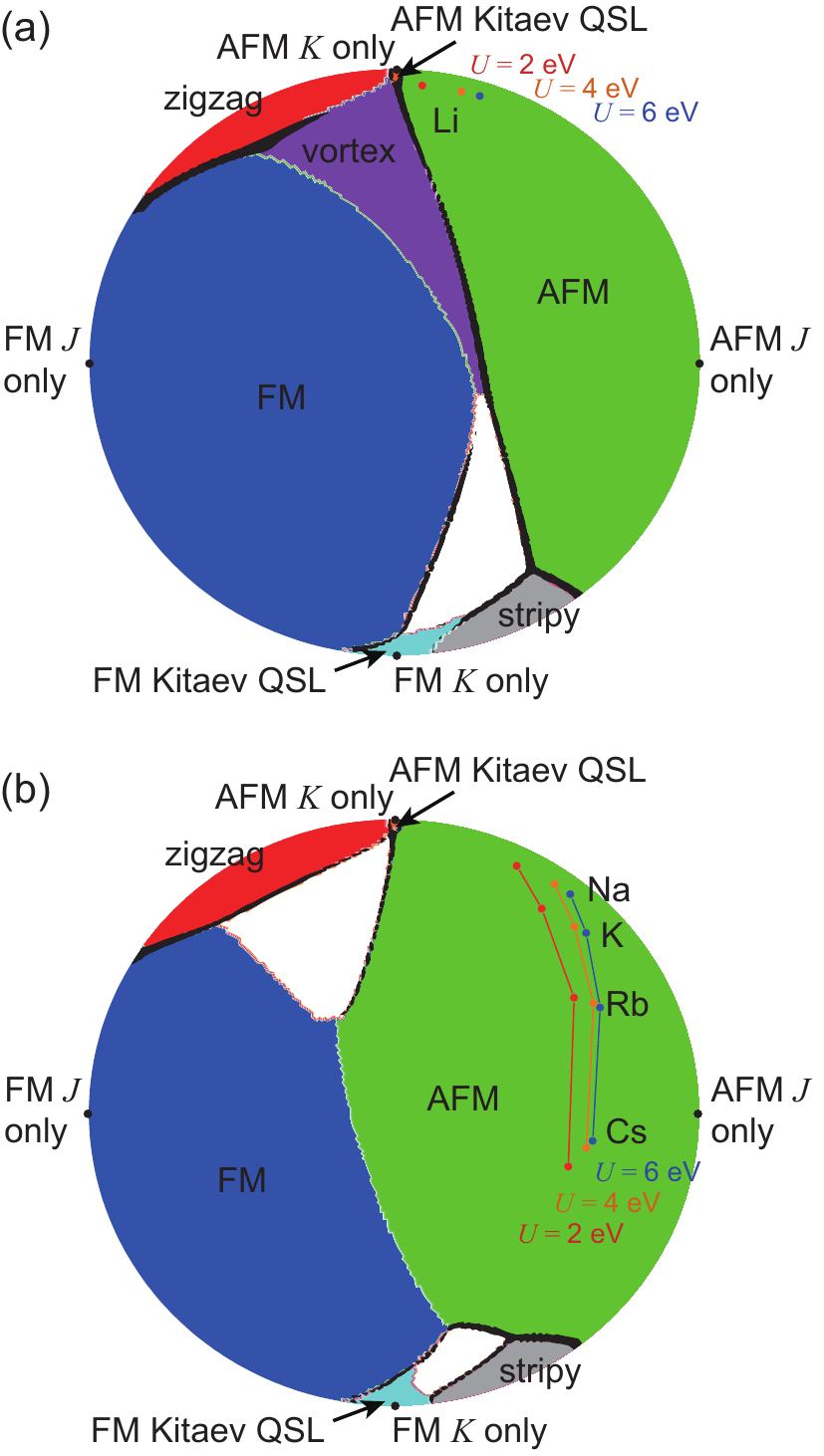}
\caption{\label{fig:f9}Phase diagrams for the $J$-$K$-$\Gamma'$ model obtained by the Lanczos exact diagonalization of a 24-site cluster. (a) and (b) show the results for $\Gamma'\leq0$ and $\Gamma'\geq0$, respectively. The Cartesian coordinates are given by $(\theta\cos\phi,\theta\sin\phi)$ and $((\pi/2-\theta)\cos\phi,(\pi/2-\theta)\sin\phi)$ for (a) and (b), respectively [see Eq.~(\ref{eq:para})], where the origins represent the $\Gamma'$-only limits ($J=K=0$). The white regions denote the regions where the spin state cannot be identified within the 24-site cluster. The red, orange, and blue dots connected by the solid lines show the evolution of the exchange coupling constants for the quasi-2D honeycomb compounds $A_2$PrO$_3$ ($A$=Li, Na, K, Rb, and Cs) at the onsite Coulomb energy $U=2$, $4$, and $6$~eV, respectively; we set the Hund's-rule coupling to $J_{\rm H}/U=0.15$, and the SOC coefficient $\lambda$ to $0.12$~eV for $A$=Li, Na, and K, and $0.11$~eV for $A$=Rb and Cs.}
\end{figure} 

Let us discuss the possible ground states for the quasi-2D honeycomb compounds $A_2$PrO$_3$, by considering the $J$-$K$-$\Gamma'$ model with the coupling constants deduced from the analyses above for $A$=K, Rb, and Cs and in the previous study for $A$=Li and Na~\cite{JA2019}. Following the previous studies of the $J$-$K$-$\Gamma$ model for the $d^5$-electron candidates~\cite{RA2014, RU2019}, we study the magnetic ground state of the $J$-$K$-$\Gamma'$ model by using the exact diagonalizations for a 24-site cluster with the Lanczos method. The results are plotted in Fig.~\ref{fig:f9} by two parameters $\theta$ and $\phi$ which are related with the coupling constants as
\begin{equation}
(J, K, \Gamma') = (\sin\theta\cos\phi, \sin\theta\sin\phi, \cos\theta).
\label{eq:para}
\end{equation}
The phase boundaries are determined by peaks in the second derivatives of the ground-state energy with respect to $\theta$ and $\phi$, and the magnetic state in each phase is identified by the spin structure factors, following the previous studies~\cite{RA2014, RU2019}. We note that the results are consistent with the previous report for the $J$-$K$-$\Gamma$ model with nonzero $\Gamma'$~\cite{RU2019}. 

As shown in Fig.~\ref{fig:f9}, large portions of the parameter space are occupied by the AFM and the FM states, extending from the trivial points in the $J$-only limits ($K=\Gamma'=0$). A classical analysis similar to Ref.~\onlinecite{CH2016B} shows that the spin moments are ordered along the $\langle111\rangle$ directions for the FM state with $\Gamma' < 0$ and the AFM state with $\Gamma' > 0$ and that the spin moments arrange in the $(111)$ plane for the FM state with $\Gamma' > 0$ and the AFM state with $\Gamma' < 0$. Meanwhile, there are small areas for the AFM Kitaev QSL and the FM Kitaev QSL states around the $K$-only limits ($J=\Gamma'=0$). Similar to the $J$-$K$-$\Gamma$ model~\cite{RA2014, RU2019}, the two QSL regions remain stable against weak $J$ and $\Gamma'$. We note that the region of the FM Kitaev QSL state for the $J$-$K$-$\Gamma'$ model is more widely spread compared to that for the $J$-$K$-$\Gamma$ model~\cite{RA2014}. The zigzag state takes place in the region with $J<0$ and $K>0$, while the stripy state appears for the opposite signs of $J$ and $K$. We note that, although Ref.~\onlinecite{RU2019} revealed two distinct zigzag patterns where spins align along the $z$ axis in the weak $\Gamma$ regime and along the $x$ and $y$ bonds, the latter is not found in the present $J$-$K$-$\Gamma'$ model. We also identify a vortex state for $\Gamma'>0$ which is similar to that found for the $J$-$K$-$\Gamma$ model~\cite{RU2019}. 

On these phase diagrams in Fig.~\ref{fig:f9}, we map out the systematic evolution of the effective coupling constants $J$, $K$, and $\Gamma'$ while changing the $A$-site cation in $A_2$PrO$_3$. The results are plotted for $U=2$, $4$, and $6$~eV with $J_{\rm H}/U=0.15$. Although all the compounds are in the AFM region, the system gets closer to the AFM Kitaev QSL region while decreasing the $A$-site ionic radii as well as the value of $U$; in particular, the $A=$Li case with $U=$2~eV is closest. Thus, our results show that the smaller $A$-site ionic radius and weaker $U$ make the system $A_2$PrO$_3$ proximate to the AFM Kitaev QSL. 

\subsection{Hyperhoneycomb magnet $\beta$-Na$_2$PrO$_3$}
\label{sec:hyperhoneycombmagnets}

\subsubsection{Lattice structure}
\label{sec:latticestructure}

\begin{figure}[t!]
\includegraphics[width=0.9\columnwidth]{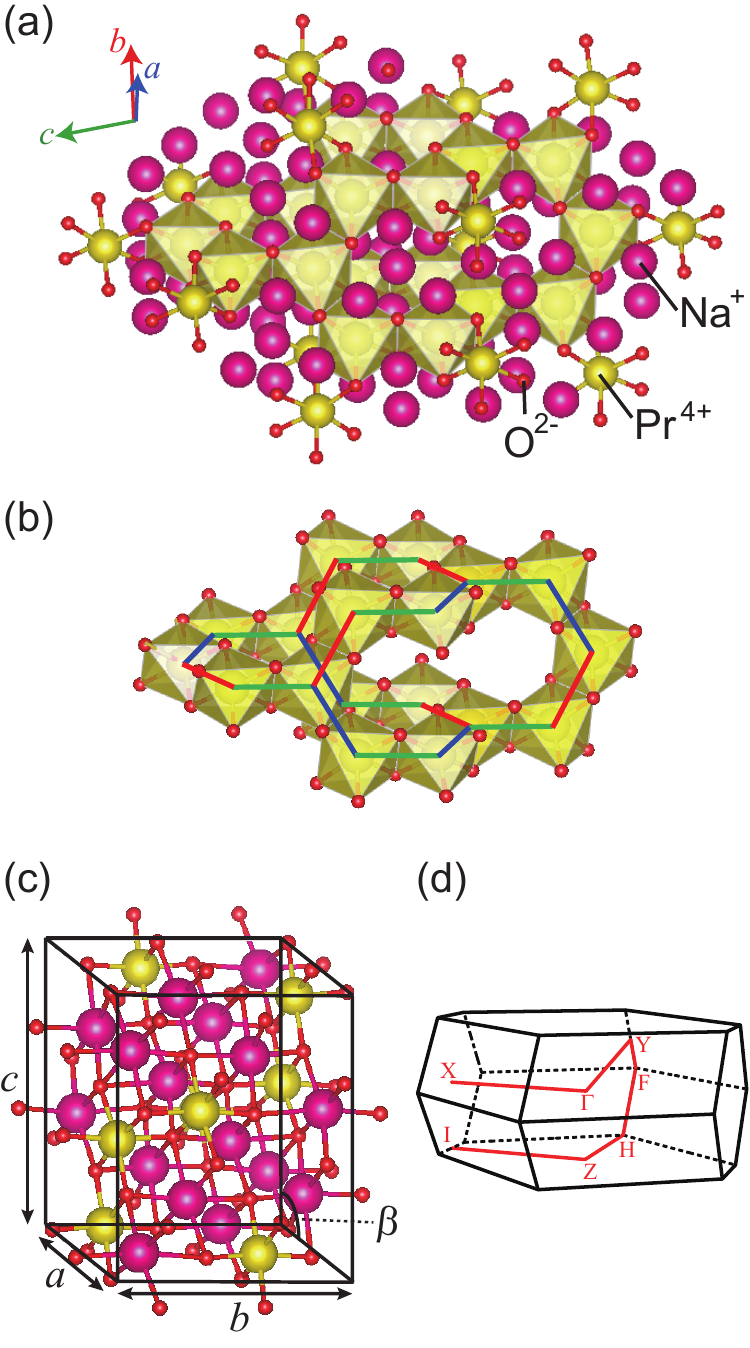}
\caption{\label{fig:f10}(a), (b), and (c) The $C2/c$ monoclinic structure of the experimentally synthesized $\beta$-Na$_2$PrO$_3$~\cite{WO1988}. The purple, yellow, and red spheres denote Na$^+$, Pr$^{4+}$, and O$^{2-}$ ions, respectively. In (a) and (b), the edge-sharing network of PrO$_6$ octahedra is partially shown. In (b), the blue, red, and green lines denote the $x$, $y$, and $z$ bonds, respectively. In (c), the black lines represent a primitive unit cell. (d) The first Brillouin zone for the monoclinic structure. The red lines represent the symmetric lines used in Fig.~\ref{fig:f11}.}  
\end{figure}

\begin{table}[b!]
\caption{\label{tab:table3}Structural parameters of the experimental structures for $\beta$-Na$_2$PrO$_3$ with $C2/c$ symmetry~\cite{WO1988}. See Fig.~\ref{fig:f10}(c) for the definitions of $a$, $b$, $c$, and $\beta$. $d_{\textrm{Pr-Pr}}$ and $\it{\theta}_{\textrm{Pr-O-Pr}}$ denote the Pr-Pr bond length and the Pr-O-Pr bond angle, respectively, for the neighboring Pr pair for the $x$, $y$, and $z$ bonds.}
\begin{ruledtabular}
\begin{tabular}{cccc}
$a$ (\si{\angstrom})&&6.7878&\\
$b$ (\si{\angstrom})&&9.7747&\\
$c$ (\si{\angstrom})&&10.806&\\
$\it{\beta}$ (deg)&&108.25&\\
\hline
&$x$ bond&$y$ bond&$z$ bond\\
\hline
$d_{\textrm{Pr-Pr}}$ (\si{\angstrom})&3.4363&3.4086&3.4400\\
$\it{\theta}_{\textrm{Pr-O-Pr}}$ (deg)&100.06&99.655&99.667\\
\end{tabular}
\end{ruledtabular}
\end{table}

\begin{figure*}[t!]
\includegraphics[width=2.0\columnwidth]{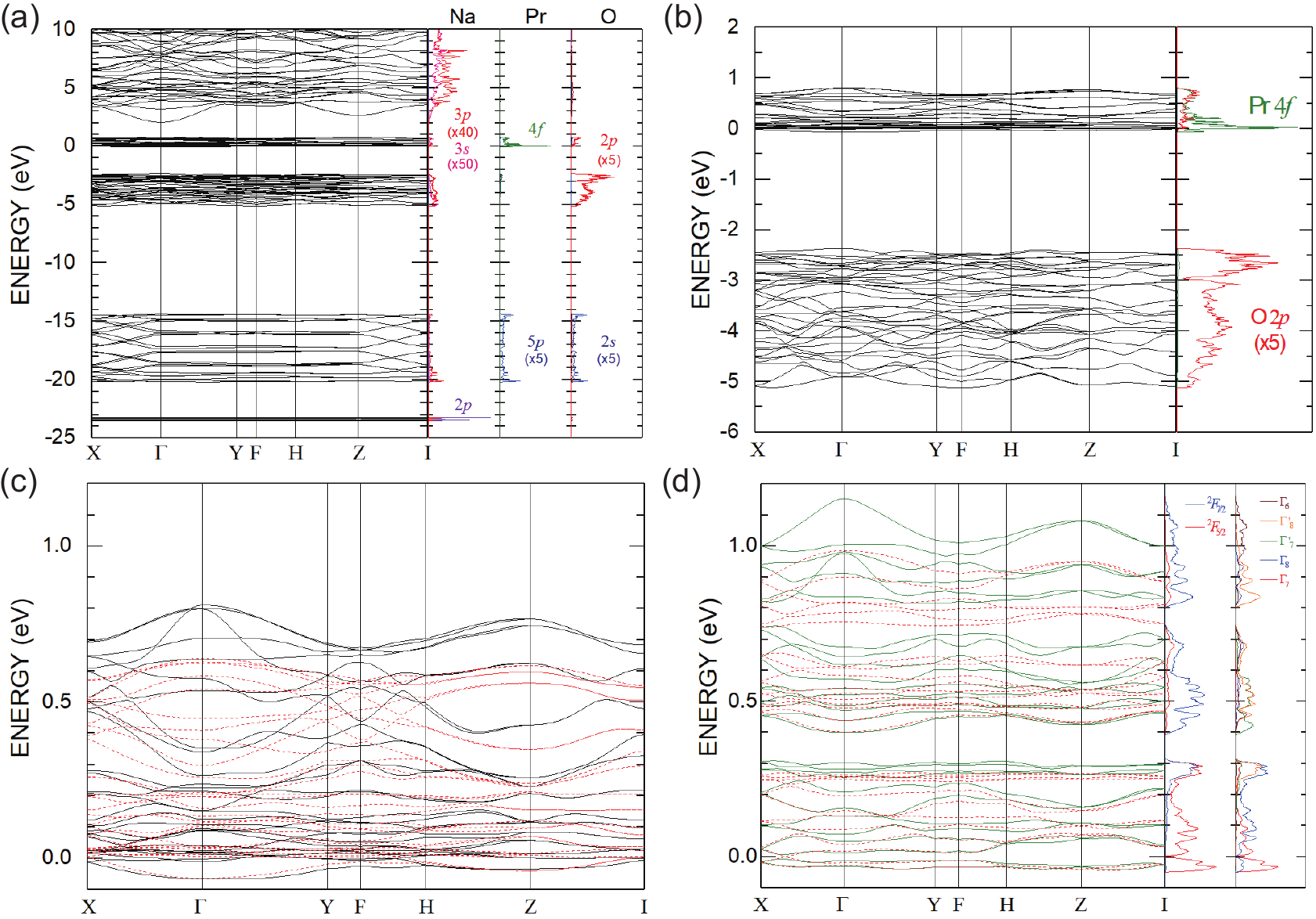}
\caption{\label{fig:f11}(a), (b), and (c) Electronic band structures for $\beta$-Na$_2$PrO$_3$ obtained by the non-relativistic {\it ab initio} calculations. The figure (a) is in the energy range from $-25$~eV to $10$~eV, (b) is from $-6$~eV to $2$~eV, and (c) is from $-0.1$~eV to $1.2$~eV. The right panels display the projected density of states to various orbitals of three atoms Na, Pr, and O in (a) and Pr $4f$ and O $2p$ orbitals in (b). In (c), the red dashed lines show the band dispersions obtained by the tight-binding calculation with nearest-neighbor transfers estimated by the MLWFs. (d) Electronic band structures for $\beta$-Na$_2$PrO$_3$ obtained by the tight-binding calculation with manually implementing the SOC Hamiltonian $\mathpzc{H}_{\textrm{SOC}}$ given by Eq.~(\ref{eq:H_SOC}) with the coefficient $\lambda$=0.12~eV~\cite{JA2019}. The green solid and red dashed lines show the band dispersions by taking into account all the transfer integrals and the nearest-neighbor transfer integrals only, estimated by the MLWFs from the non-relativistic scheme, respectively. The right panel displays the projected density of states to the $^2F_{5/2}$, $^2F_{7/2}$, $\Gamma_7$, $\Gamma_8$, $\Gamma_7'$, $\Gamma_8'$, and $\Gamma_6$ manifolds in the Pr $4f$ states. The Fermi level is set to zero.} 
\end{figure*} 

Table~\ref{tab:table3} summarizes the experimental structural parameters for $\beta$-Na$_2$PrO$_3$ with $C2/c$ symmetry~\cite{WO1988}. The experimental structure is the 3D hyperhoneycomb structure with edge-sharing PrO$_6$ octahedra, as shown in Fig.~\ref{fig:f10}. We note that the local structures indicated by $d_{\textrm{Pr-Pr}}$ and $\it{\theta}_{\textrm{Pr-O-Pr}}$ are similar to those for the quasi-2D honeycomb case of Na$_2$PrO$_3$~\cite{JA2019}.

In terms of the space group, the hyperhoneycomb structure composed of edge-sharing octahedra is seen not only in this monoclinic crystal with $C2/c$ symmetry but also in an orthorhombic crystal with $Fddd$ symmetry, as in $\beta$-Li$_2$IrO$_3$~\cite{TA2015}. The point group $D_{2h}$ of the $Fddd$ symmetry gives a $C_2$ axis that penetrates the center of the unit cell in the $[110]$ direction, other two perpendicular $C_2$ axes parallel to the $[001]$ and $[1\overline{1}0]$ directions, and the $(110)$ mirror plane. The mirror plane makes the $x$ and $y$ bonds equivalent. Meanwhile, the $C2/c$ symmetry in the present material $\beta$-Na$_2$PrO$_3$ lacks such mirror symmetry, which makes the $x$ and $y$ bonds inequivalent, as shown in Table~\ref{tab:table3}. We note that $d_{\textrm{Pr-Pr}}$ is shortest for the $y$ bond.

\begin{table}[t]
\centering
\caption{\label{tab:table4}Nearest-neighbor transfer integrals $\tilde{t}_{iu,i'v,\sigma}$ ($\sigma$=$\pm$) on a $\mu$($=x$, $y$, and $z$) bond for $\beta$-Na$_2$PrO$_3$; $u$ is in the row and $v$ is in the column. The unit is in meV. The upper-right half of the table is omitted as the matrix is Hermite conjugate.}
\begin{ruledtabular}
\begin{tabular}{cccccccc}
$\mu=x$&$\eta$&$\zeta$&$\xi$&$A$&$\beta$&$\gamma$&$\alpha$\\
\hline
$\eta$&$17.5$& & & & & & \\
$\zeta$&$-1.73$&$17.5$& & & & & \\
$\xi$&$0.87$&$0.87$&$-79.6$& & & & \\
$A$&$-1.22$&$1.22$&$1.74$&$-36.1$& & & \\
$\beta$&$-69.7$&$-8.64$&$-3.82$&$-4.52$&$132$& & \\
$\gamma$&$8.64$&$69.7$&$3.82$&$-4.52$&$-39.3$&$132$& \\
$\alpha$&$-0.96$&$0.96$&$-1.65$&$-16.1$&$-3.56$&$-3.56$&$43.7$\\ 
\hline \hline
$\mu=y$&$\zeta$&$\xi$&$\eta$&$A$&$\gamma$&$\alpha$&$\beta$\\
\hline
$\zeta$&$18.7$& & & & & & \\
$\xi$&$3.38$&$18.7$& & & & & \\
$\eta$&$0.46$&$0.46$&$-80.8$& & & & \\
$A$&$-0.95$&$0.95$&$-2.29$&$-38.3$& & & \\
$\gamma$&$-71.8$&$9.07$&$1.91$&$3.16$&$136$& & \\
$\alpha$&$-9.07$&$71.8$&$-1.91$&$3.16$&$-38.6$&$136$& \\
$\beta$&$1.16$&$-1.16$&$-0.99$&$-16.5$&$-2.83$&$-2.83$&$44.1$\\ 
\hline \hline
$\mu=z$&$\xi$&$\eta$&$\zeta$&$A$&$\alpha$&$\beta$&$\gamma$\\
\hline
$\xi$&$18.1$& & & & & & \\
$\eta$&$-4.35$&$18.1$& & & & & \\
$\zeta$&$3.41$&$3.41$&$-77.0$& & & & \\
$A$&$1.61$&$-1.61$&$2.08$&$-36.7$& & & \\
$\alpha$&$-68.6$&$-6.15$&$6.86$&$1.65$&$130$& & \\
$\beta$&$6.15$&$68.6$&$-6.86$&$1.65$&$-38.9$&$130$& \\
$\gamma$&$2.79$&$-2.79$&$-1.14$&$-17.1$&$-5.01$&$-5.01$&$41.9$\\ 
\end{tabular}
\end{ruledtabular}
\end{table}

\subsubsection{Electronic structure}
\label{sec:electronicstructure2}

Figures~\ref{fig:f11}(a), \ref{fig:f11}(b), and \ref{fig:f11}(c) display the electronic band structures and the projected density of states for the nonmagnetic state of $\beta$-Na$_2$PrO$_3$ obtained by the non-relativistic {\it ab-initio} calculations for the experimental lattice structure. The overall feature is similar to the honeycomb cases in Sec.~\ref{sec:electronicstructure}. The Pr $4f$ bands are well isolated from the other bands, locating around the Fermi level set to zero, and the bandwidth is narrow $\simeq 0.9$~eV [see Fig.~\ref{fig:f11}(c)]. Figure~\ref{fig:f11}(d) shows the band structure obtained for the tight-binding Hamiltonian constructed from the MLWF analysis for the non-relativistic band structures. As shown in this figure, when the SOC term $\mathpzc{H}_{\textrm{SOC}}$ given by Eq.~(\ref{eq:H_SOC}) with the coefficient $\lambda$=0.12~eV~\cite{JA2019} is manually implemented in the tight-binding Hamiltonian, the bandwidth of the Pr $4f$ bands is widened to $\simeq 1.3$eV. In addition, the SOC splits the $4f$ bands into the bands originating from the $^2F_{5/2}$ sextet (below $0.3$~eV) and those from the $^2F_{7/2}$ octet (above $0.3$~eV). The further decomposition of the projected density of states into the multiplets given by the OCF as represented by Fig.~\ref{fig:f1} finds that the $^2F_{5/2}$ bands and the $^2F_{7/2}$ bands are split into the $\Gamma_7$ doublet and the $\Gamma_8$ quartet and into $\Gamma_7'$ doublet, $\Gamma_8'$ quartet, and $\Gamma_6'$ doublet, respectively.

In the $4f^1$ state, the two lowest-energy shallow bands below the Fermi level (double degenerate each), which predominantly originate from the $\Gamma_7$ doublet split from the $^2F_{5/2}$ sextet, are occupied (note that the unit cell includes four Pr cations). The band gap is estimated as $\simeq 9$~meV, where the spin-orbit Mott insulator would be realized by the Coulomb interactions. In Fig.~\ref{fig:f11}(c), we show that the tight-binding band structure with the transfer integrals between nearest-neighbor Pr cations estimated from the MLWF analysis (see the next section) well reproduces the {\it ab initio} results, especially for the low-energy bands. Moreover, as shown in Fig.~\ref{fig:f11}(d), the band structure obtained only by the nearest-neighbor transfer integrals from the MLWF analysis well reproduces that by all the further-neighbor transfer integrals even when the SOC is implemented manually. Based on these observations, in Sec.~\ref{sec:couplings2}, we construct effective models for the $\Gamma_7$ pseudospins in Eq.~(\ref{eq:G7}) by taking into account only the nearest-neighbor transfer integrals.

\subsubsection{Transfer integrals}
\label{sec:transferintegrals}

Performing the MLWF analyses on the non-relativistic {\it ab initio} band structures, we estimate the transfer integrals between the Pr cations. The results for nearest-neighbor pairs on the three bonds are presented in Table~\ref{tab:table4}. We note that the values are similar to the case of the quasi-2D honeycomb compound Na$_2$PrO$_3$~\cite{JA2019}. The two transfer integrals $\tilde{t}_{i\xi,i'\alpha,\sigma}$ and $\tilde{t}_{i\zeta,i'\zeta,\sigma}$, which arise mainly from the indirect hopping processes, are comparatively large among the 11 types [Figs.~\ref{fig:f4}(a) and \ref{fig:f4}(b)], similarly to the honeycomb case in Sec.~\ref{sec:TandSOC}. These two give relevant contributions to the effective pseudospin Hamiltonian derived in the next section.

\subsubsection{Effective exchange couplings}
\label{sec:couplings2}

\begin{figure}[t!]
\includegraphics[width=0.85\columnwidth]{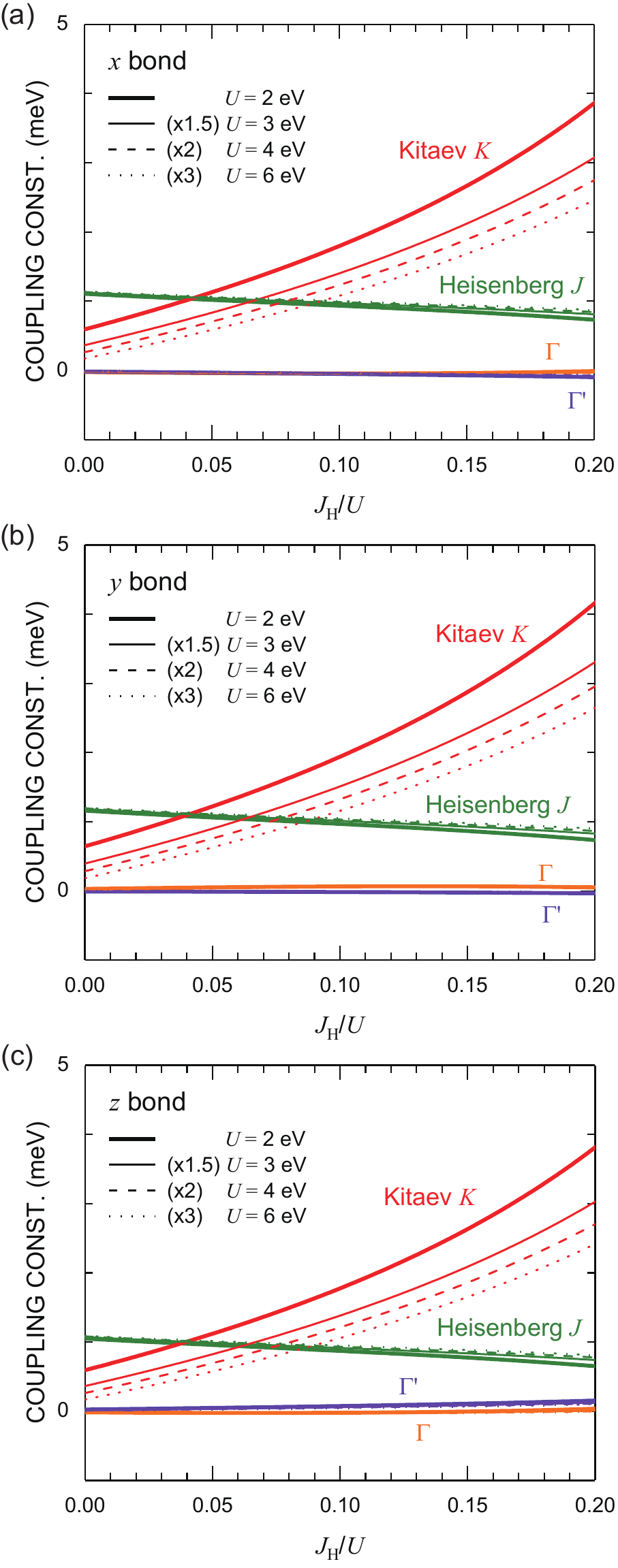}
\caption{\label{fig:f12}Coupling constants in the effective pseudospin Hamiltonian in Eq.~(\ref{eq:Heff}) for $\beta$-Na$_2$PrO$_3$ on the (a) $x$, (b) $y$, and (c) $z$ bond as functions of the Hund's-rule coupling $J_{\rm H}$. Notations are the same as in Fig.~\ref{fig:f5}. The SOC coefficient $\lambda$ is set to $0.12$~eV.}
\end{figure} 

Following the procedure of the perturbation expansion in Sec.~\ref{subsec:perturbation}, we estimate the coupling constants in the effective pseudospin Hamiltonian for $\beta$-Na$_2$PrO$_3$. The results are plotted in Fig.~\ref{fig:f12}. We take the same parameter ranges of $U=2$ -- $6$~eV and $J_{\rm H}/U=0.0$ -- $0.2$, in addition to $\lambda=0.12$~eV, as in Sec.~\ref{sec:couplings2}.

As in the quasi-2D honeycomb case~\cite{JA2019}, the AFM Kitaev coupling $K$ is most dominant for almost the entire parameter region, while the AFM $J$ is subdominant and both $\Gamma$ and $\Gamma'$ are negligibly small. The values are similar to those in the quasi-2D case~\cite{JA2019}. Our results suggest that the experimentally synthesized material $\beta$-Na$_2$PrO$_3$ is well described by the $J$-$K$ model with dominant AFM $K$, as in the quasi-2D case. We note that the amplitudes of $J$ and $K$ for the $y$ bond are comparatively larger than the other bonds, owing to the shortest bond length $d_{\rm Pr-Pr}$. Although the antisymmetric Dzyaloshinskii-Moriya interactions can be present in this structure, they are found to be negligible for all the bonds, less than $10^{-6}$~meV. We also performed the decompositions of the contributions into different perturbation processes as in Sec.~\ref{sec:couplings}, and found basically the same results. 

Similar exchange coupling constants to the quasi-2D case suggest that the ground state of $\beta$-Na$_2$PrO$_3$ is the AFM state located in the vicinity of the AFM Kitaev QSL. We note that the magnetic phase diagram for the 3D $J$-$K$-$\Gamma$ model was calculated at the classical level~\cite{LE2015, KR2019}, which appears to support our conclusion.

\section{Summary} 
\label{sec:summary}

In summary, we have systematically investigated the possible realization of Kitaev-type bond dependent interactions in $A_2$PrO$_3$ ($A=$ alkali metal) with the quasi-2D honeycomb and 3D hyperhoneycomb structures composed of edge-sharing PrO$_6$ octahedra. In these compounds, under the strong spin-orbit coupling and the octahedral crystalline electric field, the lowest-energy multiplet for the $4f^1$ electronic state of Pr$^{4+}$ cations is expected to be the $\Gamma_7$ doublet, which is described by a pseudospin with the effective angular momentum $j_{\rm eff}=1/2$. By using the {\it ab initio} calculations of the electronic band structure, we confirmed that this picture holds for all the compounds and the $\Gamma_7$ state comprises a half-filled band to be a spin-orbit Mott insulator under strong electron correlations. By constructing the multiorbital Hubbard Hamiltonian from the maximally-localized Wannier analysis and performing the perturbation expansion with respect to the electron hopping, we estimated the exchange coupling constants in the effective pseudospin Hamiltonian. 
 
In the quasi-2D case, we have studied the compounds with $A$=K, Rb, and Cs in addition to $A$=Li and Na in the previous study~\cite{JA2019}, and discussed the systematic evolution of the exchange coupling constants. We found that the low-energy magnetic properties of these compounds can be well described by the effective pseudospin Hamiltonian with the isotropic Heisenberg interaction $J$, the anisotropic Kitaev interaction $K$, and the symmetric off-diagonal interaction $\Gamma'$. These three coupling constants evolve systematically with the $A$-site substitution: (i) $J$ is dominantly AFM and does not show a drastic change, (ii) $K$ is dominantly AFM but decreases for larger $A$-site ionic radii, and finally turns into FM for $A$=Cs, and (iii) $\Gamma'$ is very small for $A$=Li and Na but increases for larger $A$-site ionic radii. Calculating the ground-state phase diagram for the $J$-$K$-$\Gamma'$ model, we showed that smaller $A$-site ionic radii make the system proximate to the AFM Kitaev quantum spin liquid, while all the compounds appear to exhibit an AFM order in the ground state. In particular, the cases with $A$=Li and Na are well described by the Heisenberg-Kitaev model ($J$ and $K$ only) with the dominant AFM Kitaev coupling $K$, as found in the previous study~\cite{JA2019}. We note that the $A$=Na case has been experimentally synthesized with a mixture of Na and Pr cations~\cite{HI2006}, while the Li case was only obtained in a different quasi-1D structure thus far~\cite{WO1987,HI2006} although our {\it ab initio} calculations suggest that the quasi-2D honeycomb structure is at least locally stable.

In the systematic study for the 2D compounds, we clarified the microscopic origin of the systematic evolution of the exchange coupling constants by carefully examining the perturbation processes and identifying the dominant hopping processes as well as the intermediate states. For smaller $A$-site ionic radii where the trigonal distortion is small, the indirect hoppings $f_\xi$-$p_x$-$f_\alpha$ (equivalent to $f_\eta$-$p_y$-$f_\beta$) and $f_\zeta$-$p_z$-$f_\zeta$ give dominant contributions to the AFM $K$, while the indirect $f_\zeta$-$p_z$-$f_\zeta$ and the direct $f_A$-$f_A$ contribute dominantly to the AFM $J$. Meanwhile, for larger $A$-site ionic radii, the increase in the trigonal distortion as well as the lattice constant weakens the dominant hoppings, but instead, it enhances the indirect $f_\zeta$-$p_z$-$f_\alpha$ (equivalent to $f_\zeta$-$p_z$-$f_\beta$) that contributes to the FM $K$, AFM $J$, and positive $\Gamma'$. As the intermediate states for the perturbation processes, the $^3H$ and $^3F$ states have the main contributions; the state $\ket{^3H_4; M_J=\pm4}$ contributes to the AFM $K$, while $\ket{^3H_4; M_J\neq\pm4}$ and $^3F$ contribute to the FM $K$. The results are distinct from those for the low-spin $d^5$ electron configuration because of the differences in the spatial anisotropy of $f$ orbitals and the atomic energy levels under the spin-orbit coupling and the crystalline electric field. 

For the 3D hyperhoneycomb case, we performed similar analyses for the experimentally-synthesized compound, $\beta$-Na$_2$PrO$_3$~\cite{WO1988}. We found that the results are similar to the 2D counterpart with $A$=Na: The effective pseudospin Hamiltonian is well described by the Heisenberg-Kitaev model. The result will stimulate material exploration of the Kitaev magnets in the series of 3D Pr-based compounds. 

We note that the energy scale of $K$ for the $f$-electron compounds is much smaller than that for $4d$ and $5d$ candidates: The former is estimated to be a few meV or less, but the latter is typically several tens of meV~\cite{CH2012, KA2014, YA2014, SA2015, BA2016, WI2016, YA2016, GL2016, BA2017, DO2017}. This is because the $f$ electrons are more localized than the $d$ electrons. Although this requires much lower temperatures to detect the interesting nature of the $f$-electron candidates, there are advantages compared to the $d$-electron cases. One is that the Kitaev coupling can be AFM, in contrast to the FM one in the existing $d$-electron candidates. This allows us to access unexplored parameter regions of the Kitaev physics. Another advantage is that parasitic magnetic orders, if any, by the non-Kitaev couplings might be destroyed by applying smaller magnetic fields because of the overall smaller energy scales. These may make possible to examine another topological phase that was recently suggested for the AFM Kitaev model in the magnetic field~\cite{ZH2017B, GO2018, NA2018, RO2019, HI2019, PA2019}.

While our analyses have been limited to the $4f^1$ case, other $f$ electron configurations may also be useful for realizing the Kitaev-type interactions. We note that there are several $f$ electron configurations that allow the lowest-energy multiplet to be the Kramers doublet~\cite{LE1962}. For instance, the lowest-energy multiplet for the $4f^5$ electron configuration, which is realized, e.g., for Sm$^{3+}$, is expected to be the $\Gamma_7$ doublet under the strong spin-orbit coupling and the octahedral crystalline electric field. In the $4f^{11}$ case, e.g., for Er$^{3+}$, while the $\Gamma_7$ doublet may compete with the $\Gamma_8$ quartet, Er$X_3$ ($X$=Br and I) was reported to show interesting magnetic properties~\cite{KR1999}, which may worth investigating the magnetic interactions from {\it ab initio} calculations like in the present study. The $4f^{13}$ case, which is the electron-hole counterpart of $4f^1$, was studied both theoretically and experimentally, as mentioned in Sec.~\ref{sec:introduction}. In this case, the expected multiplet is the $\Gamma_6$ doublet, but the competition with the $\Gamma_7'$ doublet or the $\Gamma_8'$ quartet may cause unusual magnetic interactions~\cite{RA2018}, which is potentially relevant to $\alpha$-YbCl$_3$~\cite{XI2019}. It is also interesting to note that a pyrochlore compound Yb$_2$Ti$_2$O$_7$ was discussed in the context of the Kitaev-type magnets~\cite{TH2017, PE2017, RA2019}. Beside the spin-orbit coupling and the crystalline electric field, the trigonal distortion, $p$-$f$ mixings, and $d$-$f$ electron repulsions may lead to a variety of multiplets~\cite{TA1985}. Thus, the $f$-electron compounds provide a fertile playground for exotic magnetism including the Kitaev-type quantum spin liquid. Systematic studies by extending our present work are left for future issues.

\begin{acknowledgments}
The authors thank T. Miyake for fruitful discussions. Y. M. thanks R. Coldea and K. Matsuhira for informative discussions. The crystal structures in Figs.~\ref{fig:f2}(a), \ref{fig:f2}(b), \ref{fig:f10}(a), \ref{fig:f10}(b), and \ref{fig:f10}(c) were visualized by \texttt{VESTA}~\cite{MO2011}. The second-order perturbation calculations were performed by using \texttt{SNEG} package~\cite{ZI2011}. The exact diagonalization with the Lanczos method was performed by using ${\mathcal H}\Phi$ package~\cite{KA2017}. Parts of the numerical calculations have been done using the facilities of the Supercomputer Center, the Institute for Solid State Physics, the University of Tokyo. This work was supported by JSPS KAKENHI Grant Nos.~16H02206 and 18K03447, JST CREST (JP-MJCR18T2), and US NSF PHY-1748958.
\end{acknowledgments}

\appendix
\section{Multiplets for the $f^1$ electron configuration}
\label{app:multiplets}

In this Appendix, we present the explicit forms of the multiplets for the $f^1$ electron configuration. As discussed in Sec.~\ref{subsec:KramersDoublet}, the 14-fold degenerate $f$-orbital manifold is split by the SOC into the $^2F_{5/2}$ sextet with $j=5/2$ and the $^2F_{7/2}$ octet with $j=7/2$. The $j=5/2$ manifold has the eigenvalue $-2\lambda$, which is described by the basis set 
\begin{widetext}
\begin{align}
\ket{j=\frac{5}{2}, j^z=\pm\frac{1}{2}} &= \frac{1}{2\sqrt{7}} (-\sqrt{5}{\rm i}c^{\dagger}_{\xi\mp} \mp \sqrt{5}c^{\dagger}_{\eta\mp} - \sqrt{3}{\rm i}c^{\dagger}_{\alpha\mp} \mp \sqrt{3}c^{\dagger}_{\beta\mp} \pm 2\sqrt{3}{\rm i}c^{\dagger}_{\gamma\pm})\ket{0} 
,
\label{eq:j5mj1}\\
\ket{j=\frac{5}{2}, j^z=\pm\frac{3}{2}} &= \frac{1}{2\sqrt{14}} (\sqrt{5}{\rm i}c^{\dagger}_{\xi\mp} \mp \sqrt{5}c^{\dagger}_{\eta\mp} \pm 2\sqrt{5}{\rm i}c^{\dagger}_{\zeta\pm} + 2\sqrt{5}c^{\dagger}_{A\pm} + \sqrt{3}{\rm i}c^{\dagger}_{\alpha\mp} \pm \sqrt{3}c^{\dagger}_{\beta\mp})\ket{0} 
,
\label{eq:j5mj3}\\
\ket{j=\frac{5}{2}, j^z=\pm\frac{5}{2}} &= \frac{1}{2\sqrt{14}} (-3{\rm i}c^{\dagger}_{\xi\mp} \pm 3c^{\dagger}_{\eta\mp} \mp 2{\rm i}c^{\dagger}_{\zeta\pm} - 2c^{\dagger}_{A\pm} + \sqrt{15}{\rm i}c^{\dagger}_{\alpha\mp} \pm \sqrt{15}c^{\dagger}_{\beta\mp})\ket{0} 
,
\label{eq:j5mj5}
\end{align}
while the $j=7/2$ manifold has the eigenvalue $3\lambda/2$, which is described by 
\begin{align}
\ket{j=\frac{7}{2}, j^z=\pm\frac{1}{2}} &= \frac{1}{4\sqrt{7}} (\sqrt{15}{\rm i}c^{\dagger}_{\xi\mp} \pm \sqrt{15}c^{\dagger}_{\eta\mp} + 3{\rm i}c^{\dagger}_{\alpha\mp} \mp 3c^{\dagger}_{\beta\mp} \pm 8{\rm i}c^{\dagger}_{\gamma\pm})\ket{0} 
,
\label{eq:j7mj1}\\
\ket{j=\frac{7}{2}, j^z=\pm\frac{3}{2}} &= \frac{1}{4\sqrt{7}} (5{\rm i}c^{\dagger}_{\xi\mp} \mp 5c^{\dagger}_{\eta\mp} \mp 4{\rm i}c^{\dagger}_{\zeta\pm} - 4c^{\dagger}_{A\pm} + \sqrt{15}{\rm i}c^{\dagger}_{\alpha\mp} \pm \sqrt{15}c^{\dagger}_{\beta\mp})\ket{0} 
,
\label{eq:j7mj3}\\
\ket{j=\frac{7}{2}, j^z=\pm\frac{5}{2}} &= \frac{1}{4\sqrt{14}} (\sqrt{6}{\rm i}c^{\dagger}_{\xi\mp} \mp \sqrt{6}c^{\dagger}_{\eta\mp} \pm 4\sqrt{6}{\rm i}c^{\dagger}_{\zeta\pm} - 4\sqrt{6}c^{\dagger}_{A\pm} - \sqrt{10}{\rm i}c^{\dagger}_{\alpha\mp} \mp \sqrt{10}c^{\dagger}_{\beta\mp})\ket{0} 
,
\label{eq:j7mj5}\\
\ket{j=\frac{7}{2}, j^z=\pm\frac{7}{2}} &= \frac{1}{4} (\sqrt{3}{\rm i}c^{\dagger}_{\xi\mp} \pm \sqrt{3}c^{\dagger}_{\eta\mp} - \sqrt{5}{\rm i}c^{\dagger}_{\alpha\mp} \pm \sqrt{5}c^{\dagger}_{\beta\mp})\ket{0} 
.
\label{eq:j7mj7}
\end{align}
\end{widetext}

These manifolds are further split by the OCF as discussed in Sec.~\ref{subsec:KramersDoublet}. We present the multiplets other than the $\Gamma_7$ doublet in Eq.~(\ref{eq:G7}). The $\Gamma_8$ quartet has the eigenvalue $120B_{40}$, which is described by 
\begin{widetext}
\begin{align}
\ket{j=\frac{5}{2}, \Gamma_{8a}; \pm} &= \frac{1}{6\sqrt{7}} (-\sqrt{15}{\rm i}c^{\dagger}_{\xi\mp} \pm \sqrt{15}c^{\dagger}_{\eta\mp} \pm 2\sqrt{15}{\rm i}c^{\dagger}_{\zeta\pm} + 9{\rm i}c^{\dagger}_{\alpha\mp} \pm 9c^{\dagger}_{\beta\mp})\ket{0} 
,
\label{eq:G8a}\\
\ket{j=\frac{5}{2}, \Gamma_{8b}; \pm} &= \frac{1}{14} (-\sqrt{35}{\rm i}c^{\dagger}_{\xi\mp} \mp \sqrt{35}c^{\dagger}_{\eta\mp} - \sqrt{21}{\rm i}c^{\dagger}_{\alpha\mp} \pm \sqrt{21}c^{\dagger}_{\beta\mp} \pm 2\sqrt{21}{\rm i}c^{\dagger}_{\gamma\pm})\ket{0} 
,
\label{eq:G8b}
\end{align}
the $\Gamma^{\prime}_7$ doublet from the $j=7/2$ manifold has the eigenvalue $-1080(B_{40}+14B_{60})$, which is described by 
\begin{align}
\ket{j=\frac{7}{2}, \Gamma^{\prime}_7; \pm} &= \frac{1}{\sqrt{7}} (-{\rm i}c^{\dagger}_{\xi\mp} \pm c^{\dagger}_{\eta\mp} \mp {\rm i}c^{\dagger}_{\zeta\pm} + 2c^{\dagger}_{A\pm})\ket{0} 
,
\label{eq:G7p}
\end{align}
the $\Gamma^{\prime}_8$ quartet has the eigenvalue $120(B_{40}+168B_{60})$, which is described by
\begin{align}
\ket{j=\frac{7}{2}, \Gamma^{\prime}_{8a}; \pm} &= \frac{1}{2\sqrt{7}} (\sqrt{3}{\rm i}c^{\dagger}_{\xi\mp} \mp \sqrt{3}c^{\dagger}_{\eta\mp} \mp 2\sqrt{3}{\rm i}c^{\dagger}_{\zeta\pm} + \sqrt{5}{\rm i}c^{\dagger}_{\alpha\mp} \pm \sqrt{5}c^{\dagger}_{\beta\mp})\ket{0} 
,
\label{eq:G8ap}\\
\ket{j=\frac{7}{2}, \Gamma^{\prime}_{8b}; \pm} &= \frac{1}{6\sqrt{7}} (-9{\rm i}c^{\dagger}_{\xi\mp} \mp 9c^{\dagger}_{\eta\mp} + \sqrt{15}{\rm i}c^{\dagger}_{\alpha\mp} \mp \sqrt{15}c^{\dagger}_{\beta\mp} \mp 2\sqrt{15}{\rm i}c^{\dagger}_{\gamma\pm})\ket{0} 
,
\label{eq:G8bp}
\end{align}
and the $\Gamma_6$ doublet has the eigenvalue $840(B_{40}-30B_{60})$, which is described by 
\begin{align}
\ket{j=\frac{7}{2}, \Gamma_{6}; \pm} &= \frac{1}{\sqrt{3}} (-{\rm i}c^{\dagger}_{\alpha\mp} \pm c^{\dagger}_{\beta\mp} \mp {\rm i}c^{\dagger}_{\gamma\pm})\ket{0} 
.
\label{eq:G6}
\end{align}
\end{widetext}

\section{Multiplets for the $f^2$ electron configuration}
\label{app:multiplets2}

In this Appendix, we present the multiplets for the $f^2$ electron configuration discussed in Sec.~\ref{subsec:perturbation}. In the Russel-Saunders scheme, the 91 multiplets are given in the form
\begin{widetext}
\begin{align}
\ket{L_{\rm tot},S_{\rm tot},J_{\rm tot},M_J} &= (-1)^{S_{\rm tot}-L_{\rm tot}-M_J}\sum_{M_L,M_S}\sqrt{2J_{\rm tot}+1} 
\begin{pmatrix} L_{\rm tot} & S_{\rm tot} & J_{\rm tot} \\ 
M_L & M_S & -M_J 
\end{pmatrix} 
\ket{L_{\rm tot},S_{\rm tot},M_L,M_S}
,
\label{eq:LSJMJ}
\\
\ket{L_{\rm tot},S_{\rm tot},M_L,M_S=\pm1} &=  (-1)^{-M_L}\sum_{m_1,m_2}\sqrt{2(2L_{\rm tot}+1)}
\begin{pmatrix}
3 & 3 & L_{\rm tot} \\
m_1& m_2 & -M_L
\end{pmatrix}
\tilde{c}^{\dagger}_{m_1\pm}\tilde{c}^{\dagger}_{m_2\pm}\ket{0}
,
\label{eq:LSMLMSpm1}
\\
\ket{L_{\rm tot},S_{\rm tot},M_L,M_S=0} &=  (-1)^{-M_L}\sum_{m_1,m_2}\sqrt{2L_{\rm tot}+1}
\begin{pmatrix}
3 & 3 & L_{\rm tot} \\
m_1 & m_2 & -M_L \\
\end{pmatrix}
(\tilde{c}^{\dagger}_{m_1+}\tilde{c}^{\dagger}_{m_2-}+\tilde{c}^{\dagger}_{m_1-}\tilde{c}^{\dagger}_{m_2+})\ket{0}
,
\label{eq:LSMLMSpm0}
\end{align}
\end{widetext}
where $L_{\rm tot}$, $S_{\rm tot}$, and $J_{\rm tot}$ denote the total orbital, spin, and angular momentum quantum numbers, respectively; $M_L$, $M_S$, and $M_J$ denote the total magnetic, secondary total spin, and secondary total angular momentum quantum numbers, respectively; $m_1$ and $m_2$ are the magnetic quantum numbers taking $-3,-2, \cdots ,3$. In these equations, the $2 \times 3$ matrices are the Wigner 3-$j$ symbol given by the Clebsch-Gordan coefficients.

For example, the state $\ket{L_{\rm tot}=5,S_{\rm tot}=1,J_{\rm tot}=4,M_J=\pm4}$, which is described as $\ket{^3H_4; M_J=\pm4}$, is given in the form
\begin{widetext}
\begin{equation}
\begin{split}
\ket{^3H_4; M_J=\pm4}
= \frac{1}{120\sqrt{22}}(
&\mp 6\sqrt{5}{\rm i}c^{\dagger}_{A\mp}c^{\dagger}_{\alpha\mp}
-6\sqrt{5}c^{\dagger}_{A\mp}c^{\dagger}_{\beta\mp}
+10\sqrt{3}c^{\dagger}_{A\mp}c^{\dagger}_{\eta\mp}
\mp 10\sqrt{3}{\rm i}c^{\dagger}_{A\mp}c^{\dagger}_{\xi\mp}
\mp 90\sqrt{5}{\rm i}c^{\dagger}_{A\pm}c^{\dagger}_{\alpha\pm}\\
& +90\sqrt{5}c^{\dagger}_{A\pm}c^{\dagger}_{\beta\pm}
+90\sqrt{3}c^{\dagger}_{A\pm}c^{\dagger}_{\eta\pm} 
\pm 90\sqrt{3}{\rm i}c^{\dagger}_{A\pm}c^{\dagger}_{\xi\pm}
+45\sqrt{3}{\rm i}c^{\dagger}_{\alpha\mp}c^{\dagger}_{\beta\pm}
-20\sqrt{3}c^{\dagger}_{\alpha\mp}c^{\dagger}_{\gamma\mp}\\
& -6\sqrt{5}c^{\dagger}_{\alpha\mp}c^{\dagger}_{\zeta\mp} 
-9\sqrt{5}{\rm i}c^{\dagger}_{\alpha\mp}c^{\dagger}_{\eta\pm}
\mp 36\sqrt{5}c^{\dagger}_{\alpha\mp}c^{\dagger}_{\xi\pm}
+45\sqrt{3}{\rm i}c^{\dagger}_{\alpha\pm}c^{\dagger}_{\beta\mp}
-90\sqrt{5}c^{\dagger}_{\alpha\pm}c^{\dagger}_{\zeta\pm}\\
& -9\sqrt{5}{\rm i}c^{\dagger}_{\alpha\pm}c^{\dagger}_{\eta\mp}
\mp 36\sqrt{5}c^{\dagger}_{\alpha\pm}c^{\dagger}_{\xi\mp}
\mp 20\sqrt{3}{\rm i}c^{\dagger}_{\beta\mp}c^{\dagger}_{\gamma\mp}
\pm 6\sqrt{5}{\rm i}c^{\dagger}_{\beta\mp}c^{\dagger}_{\zeta\mp} 
\pm 36\sqrt{5}c^{\dagger}_{\beta\mp}c^{\dagger}_{\eta\pm}\\
& -9\sqrt{5}{\rm i}c^{\dagger}_{\beta\mp}c^{\dagger}_{\xi\pm}
\mp 90\sqrt{5}{\rm i}c^{\dagger}_{\beta\pm}c^{\dagger}_{\zeta\pm}
\pm 36\sqrt{5}c^{\dagger}_{\beta\pm}c^{\dagger}_{\eta\mp} 
-9\sqrt{5}{\rm i}c^{\dagger}_{\beta\pm}c^{\dagger}_{\xi\mp}
\pm 12\sqrt{5}{\rm i}c^{\dagger}_{\gamma\mp}c^{\dagger}_{\eta\mp}\\
& -12\sqrt{5}c^{\dagger}_{\gamma\mp}c^{\dagger}_{\xi\mp}
\pm 10\sqrt{3}{\rm i}c^{\dagger}_{\zeta\mp}c^{\dagger}_{\eta\mp} 
+10\sqrt{3}c^{\dagger}_{\zeta\mp}c^{\dagger}_{\xi\mp}
\pm 90\sqrt{3}{\rm i}c^{\dagger}_{\zeta\pm}c^{\dagger}_{\eta\pm}
-90\sqrt{3}c^{\dagger}_{\zeta\pm}c^{\dagger}_{\xi\pm}\\
& -45\sqrt{3}{\rm i}c^{\dagger}_{\eta\mp}c^{\dagger}_{\xi\pm} 
-45\sqrt{3}{\rm i}c^{\dagger}_{\eta\pm}c^{\dagger}_{\xi\mp}
)\ket{0}.
\end{split}
\label{eq:3H4MJpm4}
\end{equation}
\end{widetext}

The energy eigenvalues of the intermediate states, $E_n$ are given as 
\begingroup
\allowdisplaybreaks
\begin{eqnarray}
E_{^3H_4}&=&F_0-25F_2-51F_4-13F_6-3\lambda,
\label{eq:E-3H4}
\\
E_{^3H_5}&=&F_0-25F_2-51F_4-13F_6-\lambda/2,
\\
E_{^3H_6}&=&F_0-25F_2-51F_4-13F_6+5\lambda/2,
\\
E_{^3F_2}&=&F_0-10F_2-33F_4-286F_6-2\lambda,
\\
E_{^3F_3}&=&F_0-10F_2-33F_4-286F_6-\lambda/2,
\\
E_{^3F_4}&=&F_0-10F_2-33F_4-286F_6+3\lambda/2,
\\
E_{^1G_4}&=&F_0-30F_2+97F_4+78F_6,
\\
E_{^1D_2}&=&F_0+19F_2-99F_4+715F_6,
\\
E_{^3P_0}&=&F_0+45F_2+33F_4-1287F_6-\lambda,
\\
E_{^3P_1}&=&F_0+45F_2+33F_4-1287F_6-\lambda/2,
\\
E_{^3P_2}&=&F_0+45F_2+33F_4-1287F_6+\lambda/2,
\\
E_{^1I_6}&=&F_0+25F_2+9F_4+F_6,
\\
E_{^1S_0}&=&F_0+60F_2+198F_4+1716F_6,
\label{eq:E-1S0}
\end{eqnarray}
\endgroup
where $F_k$ are given by the Slater-Condon parameters in Eq.~(\ref{eq:H_int}) as $F_0=F^0$, $F_2=F^2/225$, $F_4=F^4/1089$, and $F_6=25F^6/184041$; we took the ratio in values $F^2:F^4:F^6=12.980:8.163:5.878$ given by the Hartree-Fock calculation for the $4f^2$ case~\cite{FR1962}. The Coulomb repulsion $U$ and the Hund's-rule coupling $J_{\rm H}$ are given in the linear combinations of $F_k$ [see Eqs.~(\ref{eq:U}) and (\ref{eq:JH})]. 


\bibliography{Kitaev_A2PrO3}

\end{document}